\documentclass[authoryear]{elsarticle}
\usepackage{setspace}   
\usepackage{rotating}
\usepackage{float}
\usepackage{color}

\newcommand{\beginsupplement}{%
        \setcounter{figure}{0}
        \renewcommand{\thefigure}{S\arabic{figure}}%
     }

\begin{document}
\begin{frontmatter}
\begin{abstract}
\noindent The human connectome represents a network map of the brain's wiring diagram and the pattern into which its connections are organized is thought to play an important role in cognitive function. The generative rules that shape the topology of the human connectome remain incompletely understood. Earlier work in model organisms has suggested that wiring rules based on geometric relationships (distance) can account for many but likely not all topological features. Here we systematically explore a family of generative models of the human connectome that yield synthetic networks designed according to different wiring rules combining geometric and a broad range of topological factors. We find that a combination of geometric constraints with a homophilic attachment mechanism can create synthetic networks that closely match many topological characteristics of individual human connectomes, including features that were not included in the optimization of the generative model itself. We use these models to investigate a lifespan dataset and show that, with age, the model parameters undergo progressive changes, suggesting a rebalancing of the generative factors underlying the connectome across the lifespan.
\end{abstract}

\title{Generative models of the human connectome}
\author{
Richard F. Betzel$^1$,
Andrea Avena-Koenigsberger$^1$,
Joaqu\'{i}n Go\~{n}i$^{1,2}$,
Ye He$^3$,
Marcel A. de Reus$^4$,
Alessandra Griffa$^{5,6}$,
Petra E. V\'{e}rtes$^7$,
Bratislav Mi\v{s}i\'{c}$^1$,
Jean-Philippe Thiran$^{5,6}$,
Patric Hagmann$^{5,6}$,
Martijn van den Heuvel$^4$,
Xi-Nian Zuo$^3$,
Edward T. Bullmore$^7$,
Olaf Sporns$^{1,2,*}\corref{corr}$
}
\address{$^1$ Indiana University, Psychological and Brain Sciences, Bloomington IN, 47405, USA}
\address{$^2$ Indiana University, Network Science Institute, Bloomington IN, 47405, USA}
\address{$^3$ Key Laboratory of Behavioral Science and Magnetic Resonance Imaging Research Center, Institute of Psychology, Chinese Academy of Sciences, Beijing, China}
\address{$^4$ Brain Center Rudolf Magnus, Department of Psychiatry, University Medical Center Utrecht, Utrecht, The Netherlands}
\address{$^5$ Department of Radiology, University Hospital Center and University of Lausanne, Lausanne, Switzerland}
\address{$^6$ Signal Processing Lab 5 (LTS5), \'{E}cole Polytechnique F\'{e}d\'{e}rale de Lausanne, Lausanne, Switzerland}
\address{$^7$ Department of Psychiatry, Behavioural and Clinical Neuroscience Institute, University of Cambridge, Cambridge, UK}

\cortext[corr]{corresponding author: \texttt{osporns@indiana.edu}}


\end{frontmatter}

\section{Introduction}
\doublespacing
The human connectome represents a network map of the brain in which regions and inter-regional connections are rendered into the nodes and edges of a graph. In this format, the connectome can be analyzed using tools from network science and graph theory \citep{bullmore2009complex, sporns2014contributions}. Network analyses of the connectome have revealed a host of attributes that are likely essential for healthy brain function, including hierarchical and multi-scale modules \citep{bassett2010efficient, betzel2013multi}, highly connected, highly central hubs  \citep{hagmann2008mapping, van2013network}, and a rich club of mutually connected, high-degree regions \citep{van2011rich}. Additionally, the connectome's topology (the pattern in which its connections are configured) is thought to play an important role in shaping task-evoked and spontaneous brain activity \citep{hermundstad2013structural, goni2014resting, misic2015cooperative}.

The connectome is an example of a physical network whose nodes and edges are embedded in Euclidean space \citep{barthelemy2011spatial}. Consequently, the formation of connections carries a material and metabolic cost that increases with connection length \citep{bullmore2012economy}. To remain within the limits of viability, the human connectome maintains disproportionally many short-range (i.e. low cost) connections. Despite the importance of conserving connection cost, previous work in model organisms has shown that wiring minimization alone cannot account for all the connectome's topological features \citep{kaiser2006nonoptimal, da2007predicting}. Rather, connectome networks strike a balance wherein the formation of costly features like hubs and rich clubs trades off with a drive to reduce the total cost of wiring.

The conditions that allow this tradeoff to emerge are the central topic of this paper, and one that we explore using generative models applied to human connectome data obtained from individual participants. In the context of complex networks, generative modeling refers to a set of approaches for creating synthetic networks with properties similar to those of real-world networks. One example among many \citep{watts1998collective, kumar2000stochastic, sole2002model, vazquez2001modeling, dall2002random, middendorf2005inferring} is the preferential attachment model \citep{barabasi1999emergence}, which generates synthetic networks with heavy-tailed degree distributions similar to those observed in many real-world socio-technical networks.

In this report we build upon and expand the tradition of generative models for brain networks by fitting many different generative models to single-subject human connectome data and comparing models in terms of their overall performance. The models we investigate combine two distinct mechanisms for network growth: 1) geometric wiring rules which influence connection formation by favoring either shorter or longer connections and 2) non-geometric rules that ignore the distance between two regions and, instead, form connections on the basis of some shared topological relationship. Some of the models we consider implement rules that mimic well-established growth mechanisms like geometric random graphs, preferential attachment, degree assortativity, and homophilic attraction. In all cases, our aim is to discover wiring rules that produce synthetic networks with properties similar to those of observed connectomes.

To this end, we tuned our models' parameters to generate realistic synthetic networks. We found that the best-fitting model was one that penalized the formation of longer connections while increasing the likelihood of forming connections between brain regions with similar connectivity profiles (homophily). We cross-validated this result, comparing synthetic and observed connectomes along measures other than those used in the optimization process and using three different datasets. Finally, we fit the optimal generative model to data from a lifespan study (with ages ranging from 7-85 years) and found that the penalty on long-distance connections weakened monotonically with age. Older subjects' connectomes were fit poorly compared to those of younger individuals, a result driven primarily by an inability to match edge length and clustering coefficient distributions. This suggests that the human connectome undergoes a characteristic reorganization across the lifespan.

\section{Methods}

\subsection{Data acquisition and processing}
A total of $N=40$ healthy participants underwent an MRI session on a 3-T Siemens Trio scanner with a 32-channel head-coil. The magnetization-prepared rapid graident-echo (MPRAGE) sequence was 1 \textit{mm} in-plane resolution and 1.2 \textit{mm} slice thickness. The DSI sequence included 128 diffusion-weighted volumes plus one reference $b_0$ volume, maximum \textit{b}-value of 8000 \textit{s} $\cdot$ \textit{mm}$^{-2}$ and $2.2 \times 2.2 \times 3.0$ \textit{mm} voxel size. The echo planar imaging (EPI) sequence was 3.3 \textit{mm} in-plane resolution and 0.3 \textit{mm} slice thickness with TR of 1920 \textit{ms}. DSI and MPRAGE data were processing using the Connectome Mapping Toolkit \citep{daducci2012connectome}. Segmentation of grey and white matter was based on MPRAGE volumes. The cerebral cortex was parcellated into $n=219$ ROIs \citep{cammoun2012mapping}, of which we retained the 108 comprising the right hemisphere. We enforced an average connectome density of $\rho \approx 10\%$, resulting in a streamline threshold of 27 streamlines (i.e. a minimum of 27 streamlines must have connected two regions for us to consider the presence of an anatomical connection). These same data have been analyzed elsewhere \citep{avena2014using, goni2014resting, betzel2013multi}.

\subsection{Generative algorithm}
In this report we generate synthetic networks using a generative model. The algorithm for producing synthetic networks is simple. Starting with a sparse seed network (62 bi-directional edges that were common across all 40 participants), edges were added one at a time over a series of steps until $M$ total connections were placed (where $M=576\pm57$ connections across subjects). At each step we allow for the possibility that any pair of unconnected nodes, $u$ and $v$, will be connected. Connections are formed probabilistically, where the relative probability of connection formation is given by:

\begin{equation}
P(u,v) = E(u,v)^\eta \times K(u,v)^\gamma
\end{equation}

In this expression $E(u,v)$ denotes the Euclidean distance between brain regions $u$ and $v$. The exponent $\eta$ controls the characteristic connection length. When $\eta < 0$, short-range connections are favored, while $\eta > 0$ increases the probability of forming longer connections. The other term, $K(u,v)$, represents an arbitrary non-geometric relationship between nodes $u$ and $v$ and the value of $\gamma$ scales its relative importance. The precise definition of $K(u,v)$ is flexible and can be varied to realize different wiring rules. For instance, setting $K(u,v)=k_u k_v$ and $\gamma > 0$ implements a variant of preferential attachment, wherein higher degree nodes are more likely to become connected. Alternative definitions can be used to implement rules such as degree assortativity (e.g. $K(u,v)= |k_u - k_v|$, where nodes with similar/dissimilar numbers of connections preferentially connect to one another) or homophily (e.g. $K(u,v) = \sum_w a_{uw}a_{wv}$ where connections form between nodes with more or fewer common neighbors). In Table \ref{tab:tab1} we show a complete list of all non-geometric wiring rules. We limit our analysis to generative models whose wiring rules include only two components, though we could accommodate more components, in principle. We impose this limit in an effort to focus on highly simple, idealized models of network growth.

To prevent cases where $P(u,v)$ is undefined (e.g. if $K(u,v)=0$ and $\gamma<0$ then $P(u,v)=\infty$, we add $\epsilon = 10^{-6}$ to each $K(u,v)$ before raising it to the power, $\gamma$. Over the course of the generative process new edges are added to the synthetic network which necessarily changes the value of $K(u,v)$. Accordingly, at each step we update $K(u,v)$ and the corresponding changes to $P(u,v)$. If, at any step, the edge $\{u,v\}$ is  added to the synthetic network, then $P(u,v)=0$ for all subsequent steps. See Figure \ref{fig:figsi14} for an illustration of the model using a toy network model.

In our model we use Euclidean distance as a proxy for the cost of the connection between brain regions $u$ and $v$. It is worth noting that there are alternative measures for quantifying the cost or spatial relatedness of node pairs, including measures derived from the network's spatial embedding \citep{friedman2015edge}. Another candidate measure of, perhaps, greater neurobiological interest is fiber length, which measures the actual curved trajectories of white-matter tracts rather than the straight-line (Euclidean) distance between brain region centroids. While Euclidean distance and fiber length are correlated with one another, there are many instances where the fiber length of a connection is much longer than what would be expected given Euclidean distance. A more detailed discussion of this topic can be found in the Appendeix (Figures \ref{fig:figsi10} and \ref{fig:figsi11}).

\subsection{Evaluating synthetic network fitness}
To assess the fitness of a synthetic network we calculated its energy, which measures how dissimilar a synthetic network is to the observed connectome. Intuitively, if the two networks have many properties in common, then the synthetic network's energy is small. Specifically, a synthetic network's energy was defined as:
\begin{equation}
E=max(KS_k,KS_c,KS_b,KS_e)
\end{equation}
where the arguments are Kolmogorov-Smirnov statistics which quantify the discrepancy between the synthetic and observed connectomes in terms of their degree ($k$), clustering ($c$), betweenness centrality ($b$), and edge length ($e$) distributions. Here, edge length refers to the Euclidean distance between the centroids of two connected brain regions. By taking the maximum of the four statistics we consider a synthetic network to be only as fit as its greatest discrepancy.

\subsection{Model optimization}
Given the generative rule and the energy measure for evaluating a model network's goodness of fit, it was important to find the parameters $\{\eta, \gamma \}$ that produced networks with the lowest possible energy values. To solve this optimization problem, we developed a simple procedure based on classical Monte Carlo methods. The procedure consisted of three stages that were repeated:
\begin{enumerate}
\item A sampling stage in which points in parameter space are selected
\item An evaluation stage, where synthetic networks are generated with the previously-selected parameter values and their energies calculated.
\item A partitioning stage, in which the entire parameter space is partitioned according to a Voronoi tessellation.
\end{enumerate}

The procedure is initialized in stage 1 by randomly sampling $N_{samp}=2000$ points from parameter space. After evaluating the energy at each point and partitioning the entire parameter space into Voronoi cells, the algorithm returns to stage 1. Rather than sample points randomly, points are now sampled from within the boundaries of Voronoi cells, where the probability of drawing a point from within any given cell is inversely proportional to that cell's energy ($P(C) \propto E_C^{-\alpha}$, where $E_C$ is the energy of Voronoi cell, $C$, and $P(C)$ is the relative probability of sampling from within that cell). This procedure ensures that points are sampled preferentially from low-energy regions of parameter space. We repeated stages 1, 2, and 3 a total of five times and varied $\alpha$ with each repetition, going from $\alpha=\{0.0,0.5,1.0,1.5,2.0\}$. Early on, the low values of $\alpha$ meant that we searched the parameter space randomly, while the larger values at later repetitions allowed us to focus in on the low energy regions. We emphasize that alternative optimization schemes could be used to minimize $E$ (e.g. simulated annealing); the approach used here was chosen because it allowed us to not only converge to good solutions, but also to explore the energy landscape.

\section{Results}

We fit generative models to the connectomes of individual participants. In the main text, we focus on 40 adults (ages 18-40 years) scanned at the Department of Radiology, University Hospital Center and University of Lausanne (CHUV), Lausanne, Switzerland. The Appendix contains results from replication cohorts of 214 and 126 participants from the Human Connectome Project (HCP)  \citep{van2012human, glasser2013minimal} and the Nathan Kline Institute, Rockland, New York (NKI) cohort \citep{nooner2012nki}, respectively. In the same Appendix we also investigate the sensitivity of our results to alternative processing streams.

\subsection{Geometric model}
It is well known that the connectome's physical embedding shapes its topology by promoting the formation of low-cost connections \citep{bullmore2012economy}. On the other hand, forming only the shortest connections produces a skewed edge length distribution lacking long-distance connections \citep{kaiser2006nonoptimal}, resulting in increased characteristic path length, thereby reducing the efficiency with which information can flow between distant brain regions. We first sought to test the extent to which cost conservation shapes the topology of the human connectome by implementing a pure geometric model (i.e. $K(u,v)=1$).

\begin{figure}[ht]
\centering
\includegraphics[width=\linewidth]{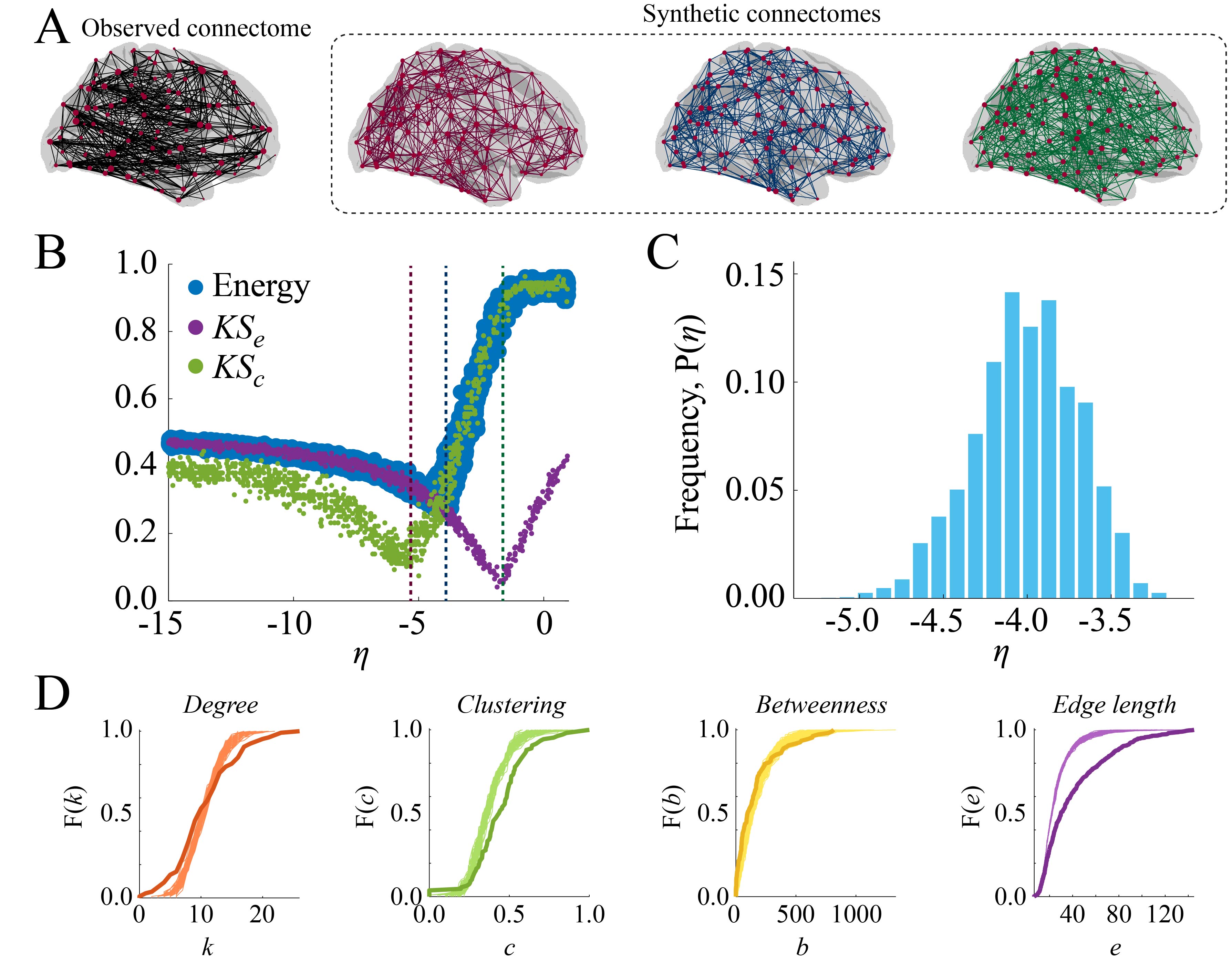}
\caption{Summary of the geometric model: (A) observed (black) and synthetic networks generated at different points in parameter space. (B) Energy landscape showing the behavior of $KS_e$, $KS_c$, and energy as a function of $\eta$. The dashed vertical lines indicate the parameter values at which the example synthetic networks were generated.  (C) Distribution of $\eta$ parameter of top 1\% lowest-energy synthetic networks aggregated across all participants. (D) Cumulative distributions of degree (orange), clustering coefficient (green), betweenness centrality (yellow), and edge length (purple) for observed connectome (darker line) and best-fitting synthetic networks (lighter lines) for a representative participant.}
\label{fig:fig1}
\end{figure}

For each participant we tuned the free parameter, $\eta$, to a range where the geometric model consistently produced synthetic networks with near-minimal energies (Figure \ref{fig:fig1}B) and analyzed the top 1\% lowest-energy synthetic networks (100 networks/participant). At this point in parameter space ($\eta=-4.01\pm0.31$; sample mean$\pm$standard error; see Figure\ref{fig:fig1}C), synthetic networks achieved an average energy of $E=0.29\pm0.02$ with KS statistics $KS_k=0.15\pm0.03$, $KS_b=0.18\pm0.04$, $KS_e=0.27\pm0.03$, and $KS_c=0.29\pm0.02$ (Figure \ref{fig:fig1}B). To contextualize these scores, we compared them to KS statistics obtained from a null generative model where connections were formed with uniform probability. We found that, with the exception of $KS_e$ ($p\approx 0.4$; Wilcoxon signed-rank test \citep{wilcoxon1945individual}), the geometric model produced significantly lower energy and smaller KS statistics (maximum $p\approx 10^{-5}$).

Interestingly, the point at which energy is minimized deviates from the respective minima of $KS_e$ and $KS_c$, demonstrating that even the-best fitting synthetic networks generated by the geometric model cannot simultaneously match observed connectomes in terms of clustering and edge length distributions. The reason for this is intuitive: A strong distance penalty is required to generate highly clustered networks, which inadvertently penalizes the formation of long-distance connections. Conversely, realistic edge length distributions arise when the distance penalty is relatively weak, at which point synthetic networks become vastly under-clustered. The energy minimum occurs at a point situated between these two extremes, trading off accuracy along one dimension with the other though never simultaneously minimizing both (Figure \ref{fig:fig1}D).

\subsection{Models driven by geometry and topology outperform pure geometric models}
\begin{figure}[ht]
\centering
\includegraphics[width=\linewidth]{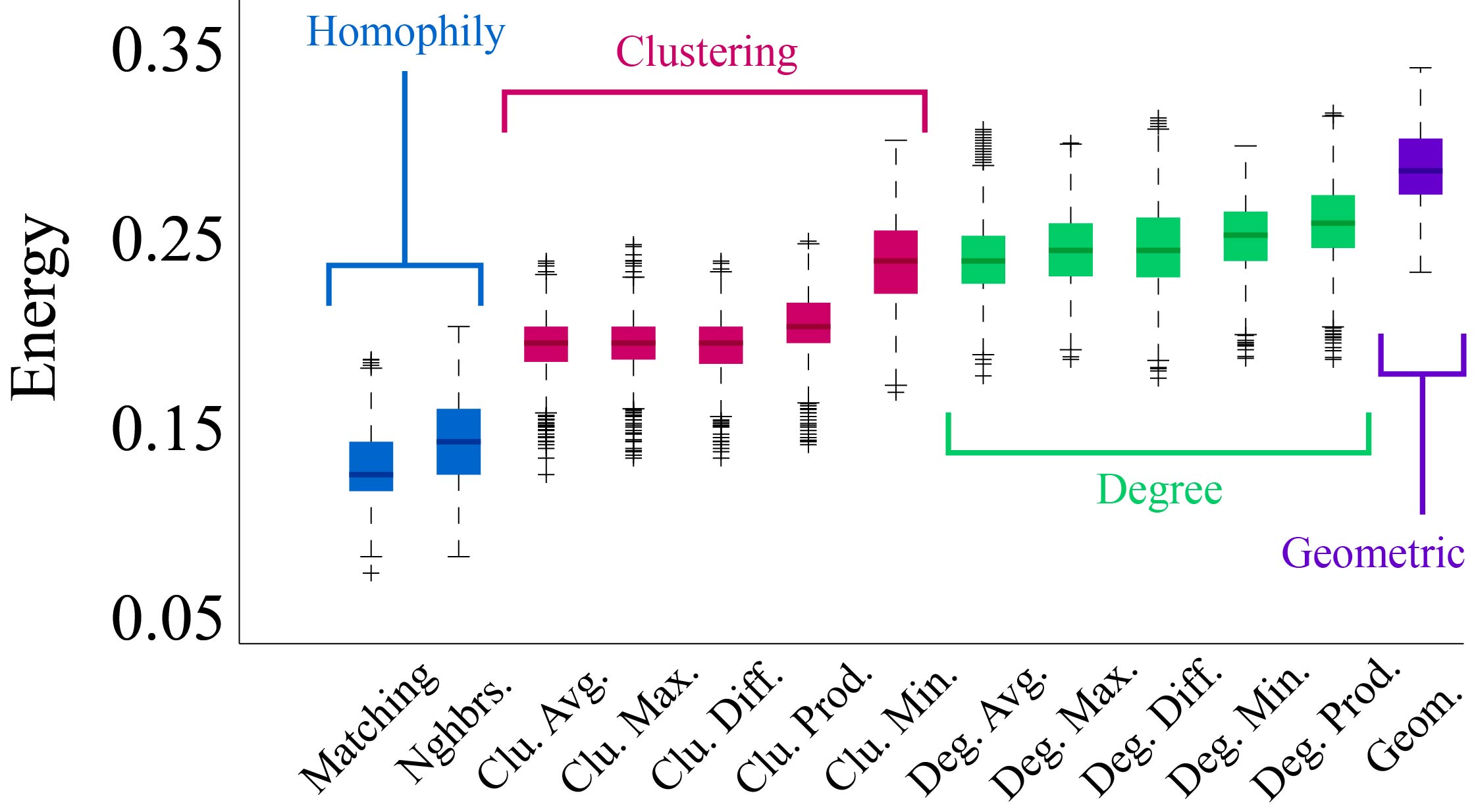}
\caption{Energy distributions across all models. Each box plot represents the top 1\% lowest energy synthetic networks generated by each model and aggregated across all participants. The color of each plot indicates the general class of the model: Homophily is shown in blue, clustering in pink, degree in green, and geometric in purple. The specific wiring rule names are shown along the x-axis.}
\label{fig:fig2}
\end{figure}

The failure of the pure geometric model to generate synthetic networks that were as clustered and contained as many long-distance connections as observed connectomes suggests that additional factors are needed as part of a realistic generative mechanism. To determine which factors were most capable in this regard we compared twelve different generative models where topological features such as degree, clustering, and homophily influenced the connection formation probabilities. As expected, due to the additional free parameter, $\gamma$, we find that all dual-factor models outperformed the pure geometric model, generating synthetic networks with significantly lower energies ($p\approx0$, see Figure \ref{fig:fig2}). Importantly, dual-factor models were stratified, with clustering-based models outperforming degree-based models, which in turn were outperformed by homophily-based models. The absolute best model incorporated a homophilic attraction mechanism in the form of the matching index (MI), which is a normalized measure of overlap in two nodes' neighborhoods. If $\Gamma_u=\{v:a_{uv}=1\}$ represents the set of node $u$'s neighbors, then the matching index is equal to:
\begin{equation}
M_{uv} = \frac{|\Gamma_{u \setminus v}\cap\Gamma_{v \setminus u}|}{|\Gamma_{u \setminus v}\cup\Gamma_{v \setminus u}|}
\end{equation}
where $\Gamma_{u \setminus v}$ is simply $\Gamma_u$ but with $v$ excluded from the set. In the event that $u$ and $v$ have perfect overlap in their neighborhoods, then $M_{uv}=1$. If the neighborhoods contain no common elements then $M_{uv}=0$.

\begin{figure}[ht]
\centering
\includegraphics[scale=0.375]{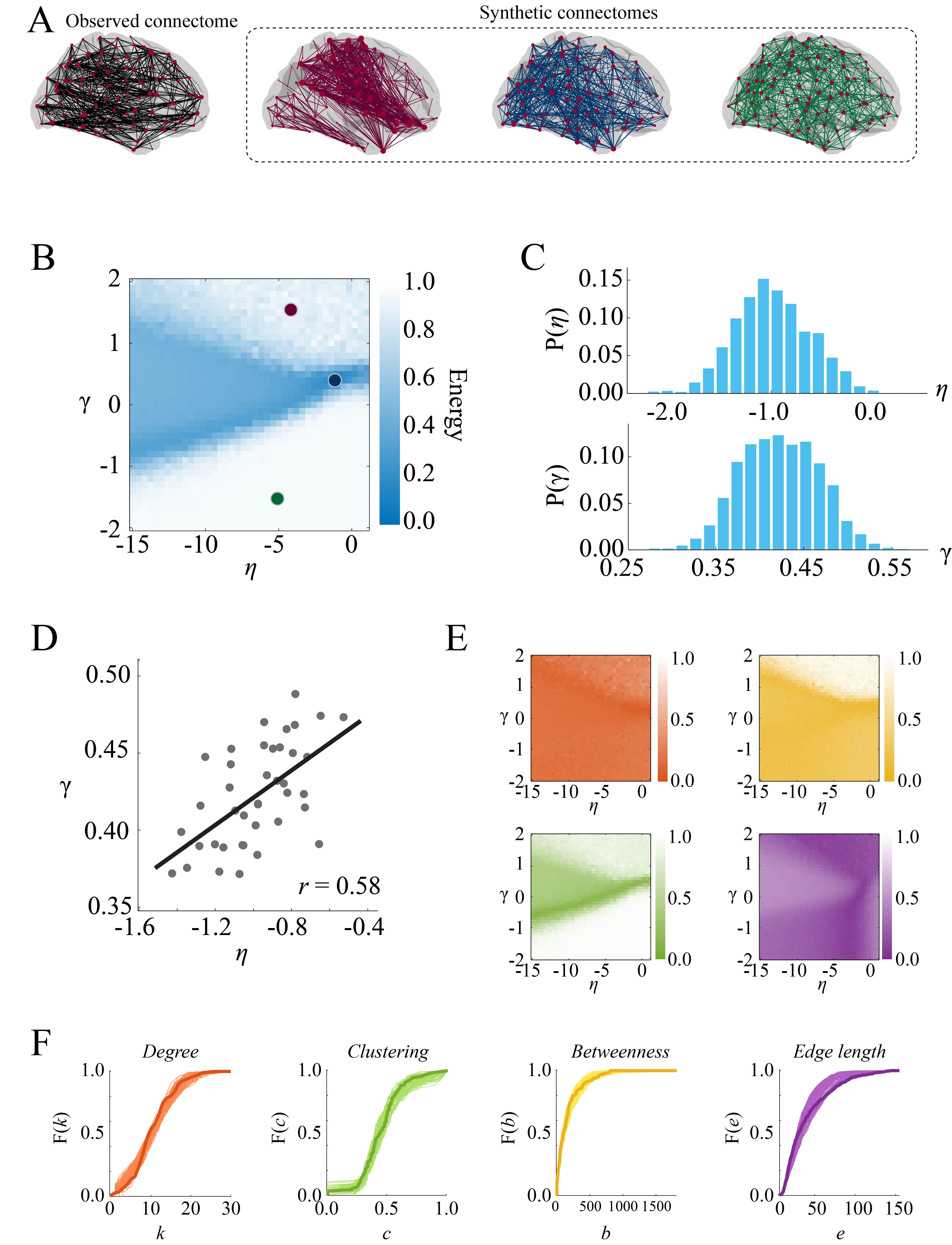}
\caption{Matching Index Model: (A) observed (black) and synthetic networks generated at different points in parameter space. (B) Energy landscape showing the points at which the example synthetic networks were generated. (C) Distribution of $\eta$ and $\gamma$ parameters of best-fitting synthetic networks aggregated across all participants. (D) Tradeoff between $\eta$ and $\gamma$. Each point represents the mean parameter values for an individual participant. Participants with larger values of $\eta$ tend to have the smallest magnitude $\gamma$ and vice versa. (E) \textit{KS} statistic landscapes for degree (orange), clustering (green), betweenness (yellow), and edge length (purple) for observed connectome and best-fitting synthetic networks for a single participant. (F) Cumulative distributions of degree (orange), clustering (green), betweenness (yellow), and edge length (purple) for observed connectome (darker line) and best-fitting synthetic networks (lighter lines) for a representative participant.}
\label{fig:fig3}
\end{figure}

Applied to the CHUV dataset, the MI model achieved an average energy of $E=0.12\pm0.02$ with parameters $\eta=-0.98\pm0.37$ and $\gamma=0.42\pm0.04$ (Figure \ref{fig:fig3}C). Together, these parameter values indicated that, like the pure geometric model, the MI model exercised a penalty against long distance connections (albeit markedly weaker than the geometric model), while increasing the probability that nodes with similar connectivity profiles would connect to one another. Interestingly, the parameters $\eta$ and $\gamma$ appear to trade off with one another (Figure \ref{fig:fig3}D), suggesting that the more an individual's connectome is shaped by geometry (large amplitude of $\eta$), the less it is shaped by non-geometric constraints and vice versa. On average, the MI model outperformed the geometric model in reducing discrepancies along all four components of the energy function: $KS_k=0.10\pm0.03$, $KS_b=0.10\pm0.02$, $KS_e=0.10\pm0.03$ and $KS_c=0.11\pm0.02$ (maximum $p$-value for all KS statistics and energy was $p\approx10^{-7}$, Wilcoxon signed-rank test). Whereas the geometric model's performance was limited primarily by mismatches in clustering and edge length, the MI model's performance was more evenly limited. The best-fitting synthetic networks had energies equal to $KS_k$, $KS_b$, $KS_c$, and $KS_e$ around 21\%, 25\%, 29\%, and 25\% of the time, respectively.

\subsection{Evaluating synthetic networks using additional measures}
Our analyses to this point consisted of tuning the parameters of generative models to ranges where the synthetic networks achieved low energy, which identified the MI model as the best fitting model. The form of the energy function, however, may be considered \textit{ad hoc}; it represents only one of many alternative ways to evaluate a synthetic network's fitness. For this reason it was important to establish that the best-fitting synthetic networks generated by the MI model matched observed connectomes across additional dimensions that were not part of the energy function used for optimization. To that end, we subjected the lowest-energy synthetic networks to a series of additional tests to determine whether they could also reproduce other properties of the human connectome.

\subsubsection{Graph theoretic measures}
\begin{figure}[ht]
\centering
\includegraphics[width=\linewidth]{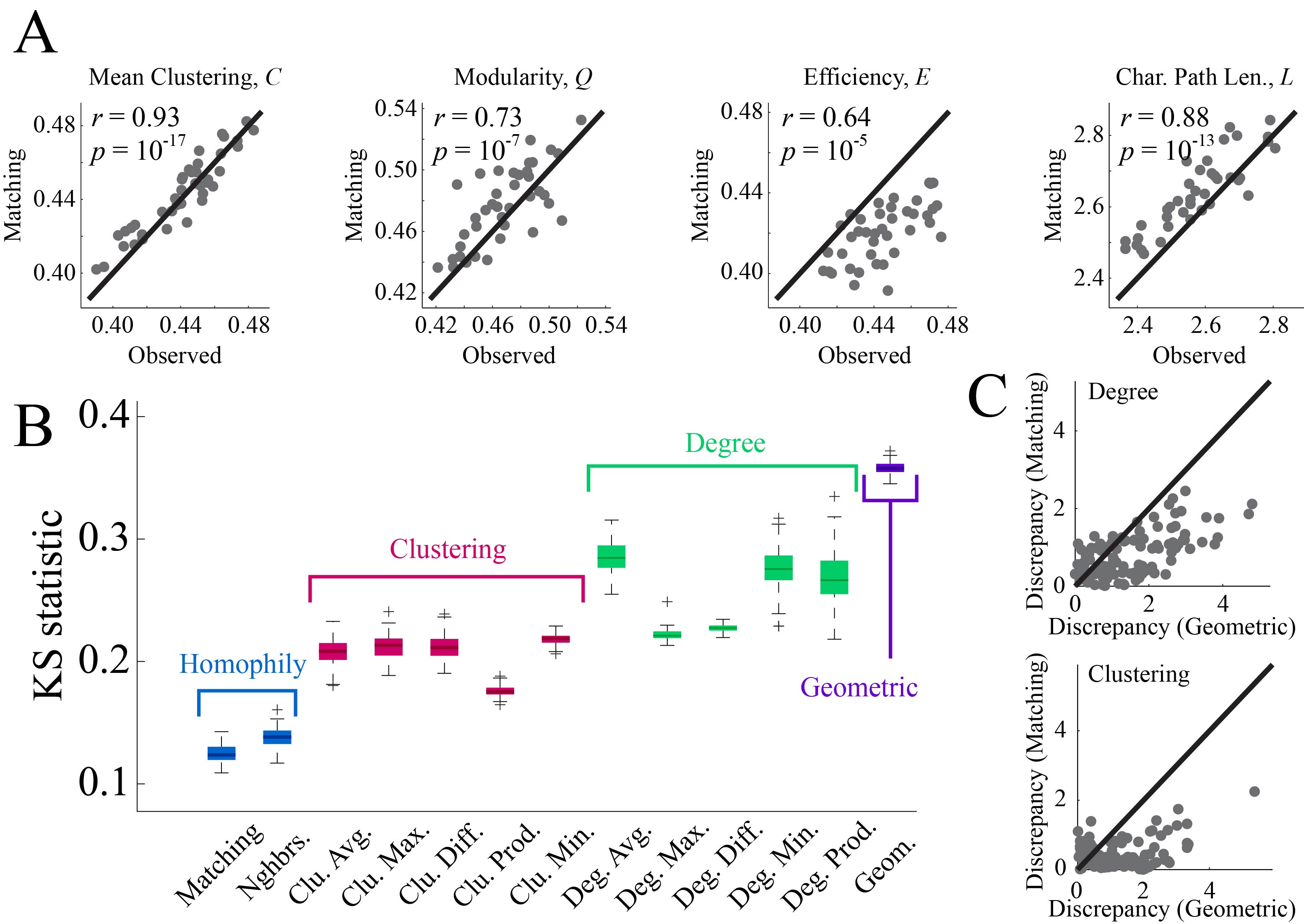}
\caption{Cross validation of the matching index model: (A) Comparison of matching index model and observed connectomes in terms of the graph-theoretic measures mean clustering coefficient, modularity, global efficiency, and characteristic path length. (B) Comparison of all models in terms of reproducing the distance-dependent degree assortativity (i.e. the propensity for high degree nodes to be linked by long-distance connections). (C) Discrepancies in degree and clustering coefficient sequences of synthetic networks generated by the matching index model and pure geometric model.}
\label{fig:fig4}
\end{figure}

The first test involved evaluating the best-fitting synthetic networks in terms of how well they matched graph-theoretical properties of observed connectomes, focusing on the measures: mean clustering coefficient ($C$), global efficiency ($E$), degree assortativity ($R_k$), modularity ($Q$), characteristic path length ($L$), and network diameter ($max[D]$) (see Appendix for descriptions of these measures). We estimated the magnitude of correlation between graph measures made on synthetic networks generated by the MI model and the same measures made on empirical networks. We found that the MI model did an excellent job reproducing the rank order of individual participants' mean clustering coefficient ($r=0.90$, $p \approx 0$), modularity ($r=0.69$, $p \approx 10^{-6}$), characteristic path length ($r=0.86$, $p \approx 10^{-12}$), and efficiency ($r=0.64$, $p \approx 10^{-5}$). Network diameter ($r=0.23$, $p=0.15$) and degree assortativity ($r=0.05$, $p=0.74$) were not well matched (Figure \ref{fig:fig4}A). It should be noted that, in general, most graph measures are not completely orthogonal to one another \citep{costa2007characterization}.

While the MI model generally reproduced the rank order of participant-level graph measures, it nonetheless systematically over-/under-estimated the values of certain measures. For instance, efficiency was, on average, smaller for synthetic networks than for empirical networks (points falling above the diagonal in Figure \ref{fig:fig4}A, third panel). The same is true for characteristic path length (over-estimated). Despite these biases, the discrepancy between empirical and synthetic networks for any of these measures was, on average, small - across participants, the mean clustering coefficient, modularity, path length, and efficiency scores of synthetic networks were always within 5.5\% of the same measure made on the corresponding observed network.

\subsubsection{Distance-dependent degree assortativity}
The human connectome features hub regions linked by long distance connections, forming rich clubs and cores. This propensity for higher-degree nodes to be linked by longer connections should be reproducible by a good generative model. To assess whether this were the case, we extracted and pooled across participants the list of all connections, the degrees of their stubs ($k_u$ and $k_v$), and length ($E(u,v)$). From these data, we estimated the three-dimensional cumulative distribution function, $F(k_\alpha,k_\beta,E(\alpha,\beta))$. At any point $\{k_\alpha,k_\beta,E(\alpha,\beta)\}$, the value of $F$ corresponded to the fraction of all connections satisfying $k_u\le k_\alpha$, $k_v \le k_\beta$, and $E(u,v) \le E(\alpha,\beta)$ ($k_u$ and $k_v$ were ordered so that $k_u \le k_v$). We constructed similar distributions for the best-fitting synthetic networks generated by each model and quantified the discrepancy between distributions with a KS statistic. In general, the rank order of models scored by this KS statistic was similar to the rank order of their energies (Figure \ref{fig:fig4}B). The MI model achieved the smallest KS statistic ($KS=0.12\pm0.01$) while the pure geometric model, on the other hand, performed the worst ($KS=0.37\pm0.01$).

\subsubsection{Local statistics}
Finally, we tested whether the best-fitting synthetic networks generated by the MI model were capable of predicting the degree and clustering coefficient sequences of the connectome. We expressed each node's empirical degree, $k_u$, and clustering coefficient, $c_u$, as z-scores by standardizing the empirical values against the distributions obtained from the best-fitting synthetic networks. Z-scores were averaged across subjects and used to quantify the discrepancy in those measures (larger scores indicated poorer fit). We compared these z-scores against scores obtained from the best-fitting synthetic networks generated by the pure geometric model in order to ascertain whether they represented an improvement in fitting local network measures (Figure \ref{fig:fig4}C). We found that, on average, the MI model produced smaller discrepancies (points below the diagonal) compared to the geometric model. Typically, the largest improvements were for nodes whose degree or clustering coefficient was mismatched the greatest by the geometric model. For some nodes, however, the geometric model actually outperformed the MI model, though the standardized scores for these nodes were, generally, rather small for both models.

\subsection{Application to human lifespan data}
\begin{figure}[ht]
\centering
\includegraphics[width=\linewidth]{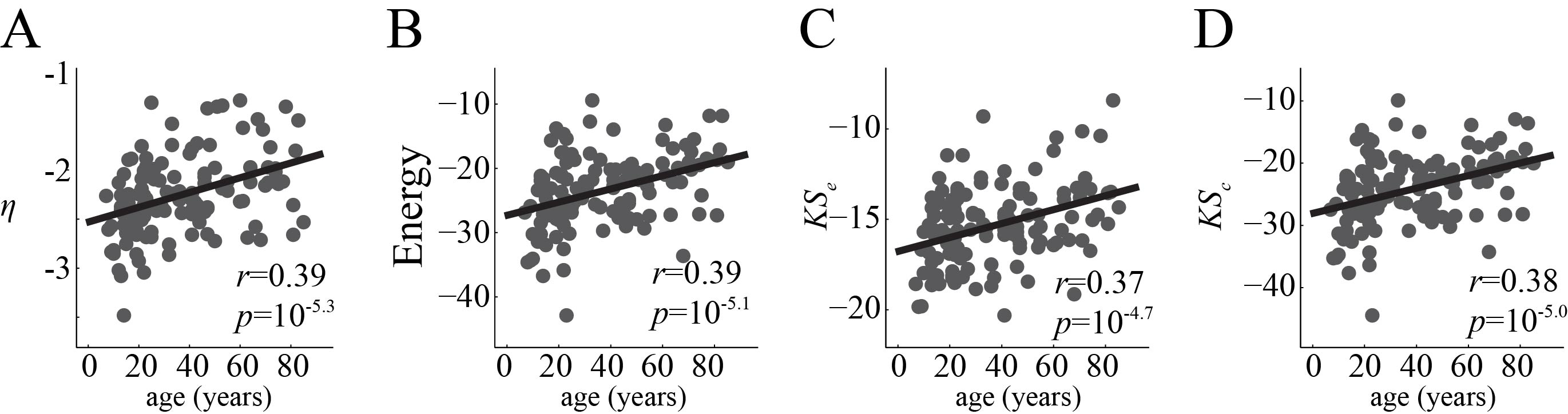}
\caption{Changes in model parameters and energy components across the lifespan: (A) The geometric parameter, $\eta$ weakens with age. (B) The average energy of each participant's best-fitting synthetic networks (z-scored against an ensemble of synthetic networks generated using a uniform wiring rule) also increases with age. (C, D) $KS_e$ and $KS_c$ increase with age, and these increases collectively drive the increase in energy.}
\label{fig:fig5}
\end{figure}

In addition to quantifying models' performances, we asked whether the parameters of the generative models captured meaningful information about individual differences in network organization. To demonstrate the utility of the network modeling approach for characterizing individual variation, we extended our analysis to the NKI dataset's $N=126$ participants, spanning a range of ages from 7-85 years. With an average network density of 10\%, a number of individual's connectomes were fractured into multiple disconnected components (71 of the 126 participants). However, the largest connected component, across all participants, included $98.5\pm0.03$ percent of all nodes, indicating that in the majority of cases the network is divided into two components: one singleton node and a component containing all other nodes. We hypothesized that age-related changes in network organization may be captured by the parameters of the generative models, $\eta$ and $\gamma$. We tested this hypothesis by first regressing out participants' intracranial volumes and mean framewise displacement from parameter values obtained from the best-fitting MI models and correlating the residuals with participant age. We also expressed energies and KS statistics as z-scores relative to a generative model in which connections were formed randomly to correct for variations in network density with age \citep{betzel2014changes, lim2015preferential}. This null model preserved only the density of connections and not degree sequence. The results of these analyses indicated that the value of $\eta$ decreased in magnitude with age ($\hat{r}_{age,\eta} = 0.39, p\approx 10^{-5.3}$), while $\gamma$ did not exhibit any significant age-related changes ($\hat{r}_{age,\gamma} = 0.07, p\approx0.45)$, which implied that the penalty on long-distance connections weakened with age. We also found that $E$, $KS_e$, and $KS_c$ all increased with age (max $p\approx10^{-4.7}$) (Figure \ref{fig:fig5}), indicating that the MI model does an increasingly poor job capturing the organization of older connectomes compared to younger connectomes.

\section{Discussion}
In this report, we tested different classes of generative models for the human connectome. Our study makes several novel contributions, by quantitatively comparing different sets of generative models, by applying these models to human connectome data, and by fitting models to networks of individual participants. We confirmed that pure geometric models of the form considered in this report cannot create synthetic networks that were both as clustered and also contained the same proportion of long-distance connections as the observed human connectome. To identify which additional factors were most capable of creating realistic networks we incorporated non-geometric information into our generative models' wiring rules. With this additional degree of freedom, the synthetic networks generated by these more complex models more accurately reproduced the connectome's clustering and edge length distributions. The best-fitting model formed connections on the basis of homophilic attraction (matching index) combined with geometric constraints. Importantly, synthetic networks generated by this model not only reproduced degree, betweenness centrality, clustering coefficient, and edge length distributions (all measures that contributed to the energy function used for optimization), but they also reproduced additional graph theoretic properties such as characteristic path length, mean clustering coefficient, global efficiency, modularity, the propensity for high-degree nodes to be connected via long-distance edges, and local node statistics such as degree and clustering coefficient sequences. We also demonstrated robustness of the matching index model, comparing it across three separate datasets totaling $N=380$ participants and finding consistent results in all cases (See Appendix). As a final demonstration of the utility of generative models, we fit the MI model to connectomes of individuals whose ages ranged from 7-85 years, showing that the distance penalty weakened with age while energy increased, an effect driven by growing discrepancies in clustering and edge length distributions.

Generative models for brain networks have been investigated before, serving as proofs of concept \citep{kaiser2004spatial, kaiser2009simple, henderson2013using, lim2015developmental, roberts2015contribution} or as investigative tools for non-human connectome data \citep{kaiser2007development, da2007predicting, nicosia2013phase, langen2015developmental}. One limitation of earlier studies was the use of composite connectivity matrices as empirical benchmarks. For example, \citet{ercsey2013predictive} and \citet{song2014spatial} proposed geometric models of an incomplete macaque connectome, where connections were based on composite tract-tracing data compiled across multiple subjects and only a subset of cortical areas. Another limitation of earlier work was the lack of model comparison. In many cases proposed generative models were only compared against random generative models \citep{ercsey2013predictive,song2014spatial} where connections were formed with uniform probability, as opposed to models incorporating more plausible generative mechanisms.

The first model we examined was the pure geometric model, which was the simplest but also, in accordance with earlier studies, performed the worst. The observation that geometry  partly explains the topology of brain networks is in line with in a large literature on wiring minimization \citep{mitchison1991neuronal, laughlin2003communication, cherniak2004global, samu2014influence}, and has been appreciated in modeling studies of both human brain networks \citep{henderson2013using, kaiser2004modelling, vertes2012simple, klimm2014resolving} and those of non-human primates \citep{kaiser2004modelling, da2007predicting} and other mammals \citep{henderson2011geometric, rubinov2015wiring}. While recent work has suggested that regional variation in certain topological properties of connectomes such as degree, clustering coefficient, and betweenness centrality, can be accounted for based on the geometry of the brain \citep{henderson2014relations}, our findings support the view that strong spatial constraints alone are insufficient for explaining all topological aspects of brain networks \citep{kaiser2006nonoptimal,bullmore2012economy}. This conclusion stands in contrast to other reports \citep{ercsey2013predictive, song2014spatial} suggesting that geometric models are the sole generative mechanism underlying the connectome's formation and evolution. Instead, we find that in order to accurately reproduce the connectome's topology our models required information about node's pairwise similarity (homophily), which agrees with earlier modeling studies of the primate connectome \citep{da2007predicting} and human functional brain networks  \citep{vertes2012simple}.

The final component of this report was an application of network modeling to human lifespan data, which revealed that geometric constraints weakened while energy and the mismatch of clustering and edge length distributions all increased with age. Collectively, these results indicate that the MI model is becoming an increasingly poor model of the connectome as participants become older. One possible explanation is that connectome patterns become increasingly random with age, making it impossible for any wiring rule to model the connectome precisely. Alternatively, it could also be the case that there are proportionally more long-range connections later in life \citep{lim2015preferential}, and therefore, with advancing age, connectomes cannot be reproduced as accurately with a wiring rule that shows preference for short-range connections. Indeed, this appears to be case; placing each participant's connections into bins (10 mm width) according to connection length and correlating bin counts with age we found that bin count was negatively correlated with age up to around 70 mm (Figure \ref{fig:figsi19}). For longer connections there was no clear relationship. Future work should investigate, in greater detail, the underpinnings of the decrease in geometric constraints.

The aim of this study was not to model the growth and development of the human connectome. Doing so would have required a more complicated model that included more system-specific detail. Instead, our models were designed to reduce a network's description length. Na\"{\i}vely, we can reconstruct a network exactly from a list of its nodes and edges. However, such a precise reconstruction may not be necessary or even desirable. Oftentimes we are more interested in a network's high-level properties (e.g. modularity, degree distribution, etc.), than the exact configuration of its connections. In such a case, a mechanism that generates synthetic networks with the approximately the same set of properties represents a much more economical (compressed) description of the network. Our models are in line with this approach, seeking a parsimonious description of the human connectome, wherein its overt complexity gets compressed into a model's wiring rule and parameters. This type of compressed description can be used toward any number of ends, including investigation of differences in individual participants. For instance, we found that some participants' connectomes were compressible (low energy) while others were not (high energy). An important question, moving forward, is whether these differences become meaningful when examining individual differences or comparing clinical and control populations, or whether they can be related to some behavioral measures across both individual and group levels.

There are a number of methodological considerations that should be discussed. First, the class of dual-term models left the definition of $K(u,v)$ up to the user. For practical reasons, we explored only twelve such rules. Even with this limited exploration, we found a great deal of stratification in terms of model performance. This leaves open the possibility that wiring rules not explored in this report could produce superior results.  For example, geometric models of forms not considered here (e.g.``connect all nodes separated by a distance less than $d_{threshold}$'', models with regional inhomogeneities (see \cite{rubinov2015wiring})), or more accurate modeling of interregional white-matter fiber lengths in place of Euclidean distance could possibly lead to improved fits. In addition, models of altogether different form could be implemented. For example, in our report we make the assumption that the formation of connections depends equally on both geometric and topological constraints. An alternative approach might be to form one subset of connections on the basis of geometry and another set of connections on the basis of topology. While enumerating of all possible wiring rules or model variations is impractical, a number of methods have been proposed that aim to discover wiring rules by evolving models themselves \citep{bailey2012automatic, menezes2014symbolic}, as opposed to proposing a model and fitting its parameters, as we did here. These approaches, we believe, warrant further attention. In any case, consideration of a broader class (or classes) of models represents an important avenue for future work.

Another methodological consideration concerns the evaluation of a synthetic network's fitness. The synthetic networks are mapped into a morphospace \citep{goni2013exploring} according to their geometric and topological properties and compared to the observed connectome along the same dimensions. Whether these properties are the most appropriate measures for network comparison is unclear. In principle, one could define alternative energy functions whose minima may not coincide with those reported here, and for which the MI model is not the best performer. Though the exploration of alternative energy functions is beyond the scope of this report, we attempted to mitigate the concern that our choice of energy function biased our results by performing a series of additional tests, the results of which indicated that the MI model consistently outperformed other models.

Another consideration relates to the combination of diffusion imaging and tractography for inferring the connectome's organization. Though diffusion imaging/tractography represents the state of the art for in vivo reconstruction of the brain's anatomical connections, these technologies are nonetheless prone to false positives and negatives \citep{thomas2014anatomical}, which could potentially affect our results. While the use of multiple atlases, independent datasets, and alternative processing streams help reduce the bias of any single processing strategy they do not completely address the issue. The shortcomings of diffusion imaging and tractography, while presently limiting, also serve to highlight the need to development new non-invasive methods for mapping the human brain.

A final consideration is related to the size of networks, the definition of nodes, and the scalability of our models. In general, how one defines a network's nodes has implications for the network properties of the resulting graph \citep{zalesky2010whole}. It is likely that the size and number of nodes factor into the performance of the models studied here. The networks analyzed in this report consisted of either $n=74$ or $n=108$ nodes, representing two different parcellations of the cortex. However, it is becoming increasingly common to model brain networks with up to thousands of nodes. Because the number of possible positions to place an edge grows as $O(n^2 )$, the space of all networks that the model could generate becomes much larger as n increases. Models with $n>>10^2$ may require stronger parametric constraints (e.g. larger magnitudes for $\eta$ or $\gamma$) or incorporating additional topological information (and an additional parameter) into a model's wiring rule. In general, the choice of how to define a network's nodes and at what scale the human connectome is best described is unclear, though future work on data-driven parcellations will surely help address this issue.

\section{Appendix}
The main text describes the results of generative models applied to a dataset of 40 participants scanned at CHUV. In this appendix we demonstrate the robustness of those results by reproducing the principal findings using alternative datasets. The additional datasets are described, briefly, below and in more detail later in this appendix. Figures S1-S9 shows model energies for each of the additional datasets, reproducing Figure 2 from the main text.

\begin{enumerate}
\item Two replication datasets (HCP and NKI) of $N=214$ and $N=126$ participants, respectively.
\item The same CHUV dataset with different levels of network density (5\% and 15\%) and defined using an alternative weighting.
\item CHUV dataset including both left/right cerebral hemispheres.
\item Composite (i.e. group averaged) CHUV, HCP, and NKI connectomes.
\item Composite CHUV dataset using fiber length in place of Euclidean distance.
\item Composite CHUV dataset using an exponential function to model geometric constraints in place of the power-law function.
\item Composite CHUV dataset using a finer cortical parcellation ($n=223$ nodes.)
\end{enumerate}

Each additional dataset is accompanied by a figure showing the energy distribution for the 100 best-fitting synthetic networks for each model type. At the end of this appendix we have also included a glossary of graph theoretic terms that appear throughout the main text.

\section*{Additional Datasets}
\subsection{Human connectome project (HCP) - See Figure S1}
The HCP data were drawn from the 215 participants made available as part of the Q3 release of the human connectome project \citep{van2012human, glasser2013minimal}. From each participant's diffusion-weighted MR images (diffusion tensor imaging; DTI), white matter fibers were reconstructed from generalized q-sampling \citep{yeh2010generalized} (GQI: allowing for the reconstruction of crossing fibers) and streamline tractography and the cortex was parcellated into 219 parcels based on a subdivision of FreeSurfers's Desikan-Killiany atlas \citep{cammoun2012mapping}. More details on the processing of these data can be found elsewhere \citep{de2014simulated}. We focused on the right hemisphere only, which consisted of $n=108$ regions. We imposed a threshold on streamline counts of 5 (i.e. a minimum of five streamlines must be present for us to consider two regions linked by a binary connection) in order to maintain an average connectome density of $\rho \approx 10\%$ across subjects. We excluded a single subject on the grounds that their total streamline count was greater than two standard deviations from the group mean, leading to a final dataset of $N=214$ participants.

\subsection{Nathan Kline Institute, Rockland, NY (NKI) - See Figure S2}
The NKI dataset consists of $N=126$ participants whose ages ranged from 7-85 years \citep{nooner2012nki}. Tractography was performed using the Connectome Computation System (CCS: http://lfcd.psych.ac.cn/ccs.html). A more detailed description of the processing pipeline was included in other reports \citep{betzel2014changes, cao2014topological, yang2014connectivity}. Unlike the HCP and CHUV datasets, the cortex was parcellated into 148 regions according to the Destrieux atlas \citep{destrieux2010automatic}. We analyzed a single hemisphere ($n=74$ regions), but instead of focusing on either the right or left, we formed a composite matrix by combining the streamline counts between homotopic pairs of regions. We, again, enforced a mean density of $\rho \approx 10\%$ by selecting a streamline threshold of 30 streamlines.

\subsection{Alternative CHUV datasets - See Figures S3-S6}

We investigated four variants of the CHUV dataset. In the main text we analyzed binary connectivity matrices (average density of $\rho \approx 10\%$) by applying a threshold to streamline counts. The first two variants were constructed in the same manner but with the threshold level chosen to maintain average densities of $\rho \approx 5\%$ and $\rho \approx 15\%$. The third variant retained a threshold of $\rho \approx 10\%$ but instead of thresholding streamline counts we thresholded "fiber density" matrices. The fiber density between nodes $u$ and $v$ is common choice for edge weights in weighted anatomical brain networks, and is defined as the number of streamlines divided by the sum of $u$ and $v$'s surface areas \citep{hagmann2008mapping, betzel2013multi, goni2014resting}. The fourth variant was constructed by thresholding streamline counts to $\rho \approx 10\%$ but included both left and right cerebral hemispheres.

\subsection{Group-average matrices - See Figures S7-S9}

In addition to single-participant modeling, we analyzed group-average connectivity matrices for all three datasets (CHUV, HCP, and NKI). Group-average matrices boost the ratio of signal to noise by emphasizing connections that are consistently expressed across subjects, thereby rendering the human connectome more clearly. The \textit{de facto} method for generating group-average matrices is to retain the supra-threshold elements of the $[n \times n]$ consistency matrix, $\mathbf{C}$, whose element $c_{uv}$ indicates the fraction of all participants in which a connection was present between nodes $u$ and $v$. The resulting matrix, however, over-expresses short-range connections, as short-range connections are more easily reconstructed and are hence the most consistent connections across subjects whereas long-range connections are more prone to error. Also, this method forces a user to choose, somewhat \textit{ad hoc}, the threshold for including a connection in the group-average matrix. Instead, we use an alternative method for generating a group-average connectomes whose edge-length distribution matches that of the typical single-participant distribution \citep{misic2015cooperative}. Briefly, this method begins by first estimating the average number of connections of a given length in a typical participant's connectome. Next, all pairs of nodes separated by a comparable distance are identified and, from among this subset, the most consistent connections are added to the group-average connectivity matrix. Repeating this process for all distances yields a representative connectome that matches, almost exactly, the typical edge length distribution, but features only the most consistently expressed edges at each connection length.

\subsection{CHUV Group-average matrix with fiber length - See Figures S10-S11}

In this report, we test the hypothesis that the human connectome emerges as a consequence of both topological and spatial constraints, which we model as power-law functions. In doing so, we assume that the material/metabolic cost of fiber tracts can be equated to Euclidean distance separating its endpoints, rather than the actual integrated length of the curved tract. The argument for doing so is threefold. First, estimates of fiber length can only be obtained for completed streamlines, meaning that no estimates exist for connections that were absent in the observed tractography data. In order to fill in the missing fiber lengths, one can resort to fiber interpolation (i.e. using the distance/fiber length relationship of existing connections to estimate the fiber length of missing connections), which necessarily introduces an additional source of uncertainty. Second, the relationship of fiber length and Euclidean distance is rather strong across our datasets: the amount of variance in fiber length accounted for by Euclidean distance was 66\%, 32\%, and 79\% for the CHUV, NKI, and HCP datasets, respectively. Lastly, a recent study used both Euclidean distance and the measured length of axons in a geometric generative model of the mouse connectome \cite{rubinov2015wiring}, finding no change in their results. For these reasons, we assert that Euclidean distance, though imperfect, is a reasonable proxy for the cost of forming a connection.

Nonetheless, we felt it prudent to test the effect of using fiber length in place of Euclidean distance in our models. To do so we first constructed an interpolated fiber length matrix. The elements of this matrix contain fiber length estimates for the hypothetical tracts linking the pair of nodes, $u$ and $v$. To obtain such estimates for nodes $u$ and $v$, we calculate the Euclidean distance, $E(u,v)$ between their respective centroids. We then consider all geometric neighbors of $u$ and all geometric neighbors of $v$, where a geometric neighbor of $u$ is any other node whose centroid is less than $E(u,v)\times \tau$ away from that of $u$. Here, we use a fixed value $\tau=0.2$. The fiber length of the hypothetical connection between nodes $u$ and $v$ is set equal to the mean fiber length of connections between $u$ and any of $u$'s geometric neighbors and $v$ and any of $v$'s geometric neighbors. If no connections exist between these subsets of nodes, then we used the $\beta$ coefficients from the Euclidean distance versus fiber length linear regression model to estimate a fiber length.

\subsection{CHUV Group-average matrix with exponential rule - See Figure S12}
In the main text we modeled the geometric wiring rule as a power-law function. However, recent work has suggested that an exponential function better captures the relationship between edge length and connection probability \citep{henderson2014relations}. To this end, we replaced the geometric power-law function in our geometric models with an exponential function and re-ran our analyses.

\subsection{CHUV Group-average matrix with finer cortical parcellation - See Figure S13}
The pipeline described in \cite{cammoun2012mapping} makes it possible to partition the cerebral cortex into parcels at several different scales or resolutions. In our main analysis, we used an intermediate scale ($n=219$ cortical parcels, with $n=108$ parcels in the right hemisphere). In this section, we repeat our analysis for a composite group-average CHUV matrix generated at the next finest scale, which has $n=223$ cortical parcels in the right hemisphere. The group-average matrix was generated using the same procedure as described earlier.

\subsection{Graph theory}
In the main text we characterize networks using a number of different graph-theoretic measures. Here we describe those measures in some detail \citep{rubinov2010complex}.

\begin{itemize}

\item \textit{Adjacency matrix}: A topology of an $n$-node network can be described by the matrix $\mathbf{A}=[a_{uv}]$. The elements $a_{uv}$ are equal to 1 if nodes $u$ and $v$ are connected and are otherwise equal to 0.

\item \textit{Node degree}: A node's degree counts the total number of connections that node makes. In an undirected network (i.e. $a_{uv} = a_{vu}$) a node's incoming and outgoing degrees are equivalent, and can be calculated as the sum across rows or columns of $\mathbf{A}$, i.e. $k_u = \sum_v a_{uv}$.

\item \textit{Network density}: A network's density, $\rho$, is equal to the fraction of existing connections out of the total number of possible connections. If the total number of connections is equal to $2m = \sum_u k_u$, then network density is equal to $\rho = \frac{2m}{n(n-1)}$.

\item \textit{Degree assortativity}: Degree assortativity measures the extent to which nodes of similar degree connect to one another. It is usually operationalized as a correlation measure, $R_k$, which measures the Pearson correlation of the stub degrees of all edges \citep{newman2002assortative}.

\item \textit{Clustering coefficient}: A node's clustering coefficient measures the density of a node's neighborhood. Phrased differently, clustering coefficient it measures the fraction of a nodes' neighbors that are connected to one another. If $t_u=\frac{1}{2} \sum_{vw} a_{uv} a_{uw} a_{vw}$ is the number of triangles (connected neighbors) surrounding node $u$, then $u$'s clustering coefficient is equal to $c_u = \frac{2 \cdot t_u}{k_u (k_u - 1)}$. The mean clustering coefficient of a network is simply the average of $c_u$ over all possible $u$.

\item \textit{Characteristic path length}: The shortest path matrix, $\mathbf{D} = [d_{uv}]$, measures the length of the shortest paths between all pairs of $u$ and $v$. The characteristic path length is the average length of all shortest paths and is calculated as $L = \sum_{u,v \ne u} \frac{d_{uv}}{n(n - 1)}$.

\item \textit{Network diameter}: A network's diameter is the longest shortest path between any two nodes, and is calculated as $max (d_{uv})$.

\item \textit{Global efficiency}: A network's efficiency is closely related to characteristic path length. Rather than calculating the average length of a shortest path, efficiency is calculated as the average length of $\frac{1}{d_{uv}}$. Specifically, a network's efficiency is calculated as $E = \sum_{u,v \ne u} \frac{d_{uv}^{-1}}{n(n-1)}$.

\item \textit{Modularity}: Many network measures describe a network's organization at the level of individual nodes or with a global summary statistic. Alternatively, it is possible to describe a network's "large-scale structure" \citep{newman2012communities} - i.e. its organization at an intermediate scale. Perhaps the most common type of large-scale structure is known as a network's community structure or a division of a network into internally dense and externally sparse modules \citep{fortunato2010community, sporns67modular}. The most popular method for identifying a network's communities and evaluating their fitness is to maximize a "modularity" function \citep{newman2004finding}:
\begin{equation}
Q=\frac{1}{2m}\sum_{uv} [a_{uv} - p_{uv}]\delta(g_u,g_v)
\end{equation}
In this expression, $g_u \in \{1,\dots,K\}$ is the community to which node $u$ is assigned, $\delta(g_u,g_v)$, is the Kronecker delta function and is equal to unity when $g_u = g_v$, and $p_uv$ indicates the expected number of connections between $u$ and $v$ under a particular null model (typically $p_{uv} = \frac{k_uk_v}{2m}$, which is the expected weight under the null model where each node's degree is preserved but connections are otherwise made randomly). In general, high quality modules produce large values of $Q$. To maximize $Q$, then, one needs to ensure that connections satisfying $(a_{uv} - p_{uv}) > 0$ fall within communities. The process of modularity maximization is computationally intractable for all but the most trivial cases, though many heuristics are available for identifying near-optimal modules. Here, we use the Louvain algorithm \citep{blondel2008fast} to produce 100 near-optimal estimates of modules.
\end{itemize}

\section{Acknowledgements}

Data were provided [in part] by the Human Connectome Project, WU-Minn Consortium (Principal Investigators: David Van Essen and Kamil Ugurbil; 1U54MH091657) funded by the 16 NIH Institutes and Centers that support the NIH Blueprint for Neuroscience Research; and by the McDonnell Center for Systems Neuroscience at Washington University.

\section{Funding information}
R.F.B. was supported by the National Science Foundation/Integrative Graduate Education and Research Traineeship Training Program in the Dynamics of Brain-Body-Environment Systems at Indiana University.
J.G. was supported by the Indiana University Network Science Institute and the Indiana Alzheimer Disease Center (NIH/NIA P30 AG10133).
A.G. was supported by the Swiss National Science Foundation (320030 130090).
P.E.V. was supported by a Bioinformatics Research Fellowship from the Medical Research Council (UK) (MR/K020706/1).
B.M. was supported by a Natural Sciences and Engineering Research Council of Canada Postdoctoral Fellowship.
P.H. was supported by the Leenaards Foundation.
M.H. was supported by VENI (451-12-001) grant of the Netherlands Organization for Scientific Research (NWO; http://www.nwo.nl) and by a Fellowship of the Brain Center Rudolf Magnus.
X.Z. was supported by the National Basic Research Program (973) of China (http://www.most.gov.cn/eng/), grant number is 2015CB35170 and the Major Joint Fund for International Cooperation and Exchange of the National Natural Science Foundation (http://www.nsfc.gov.cn/publish/portal1/), grant number is 81220108014.
E.T.B. was supported by a grant from the Behavioural \& Clinical Neuroscience Institute (University of Cambridge) and by the Wellcome Trust and the Medical Research Council.
O.S. was supported by the J.S. McDonnell Foundation (220020387) and the National Science Foundation (1212778). 

\section{Author contributions statement}

R.F.B., P.E.V., E.T.B., O.S. conceived the experiment(s),  R.F.B., A.A-K., B.M., J.G., and O.S. conducted the experiment(s),  R.F.B., A.A-K., B.M., J.G., and O.S. analyzed the results, Y.H., M.A.R., A.G., J-P.T., P.H., M.H., and X-N.Z. contributed data.  All authors reviewed the manuscript. 

\section{Additional information}
ETB is employed half-time by the University of Cambridge and half-time by GlaxoSmithKline (GSK); he holds stock in GSK. The authors declare no other competing financial interests.

\clearpage

\section*{References}
\bibliography{biblio}

\beginsupplement
\begin{figure}[ht]
\centering
\includegraphics[width=\linewidth]{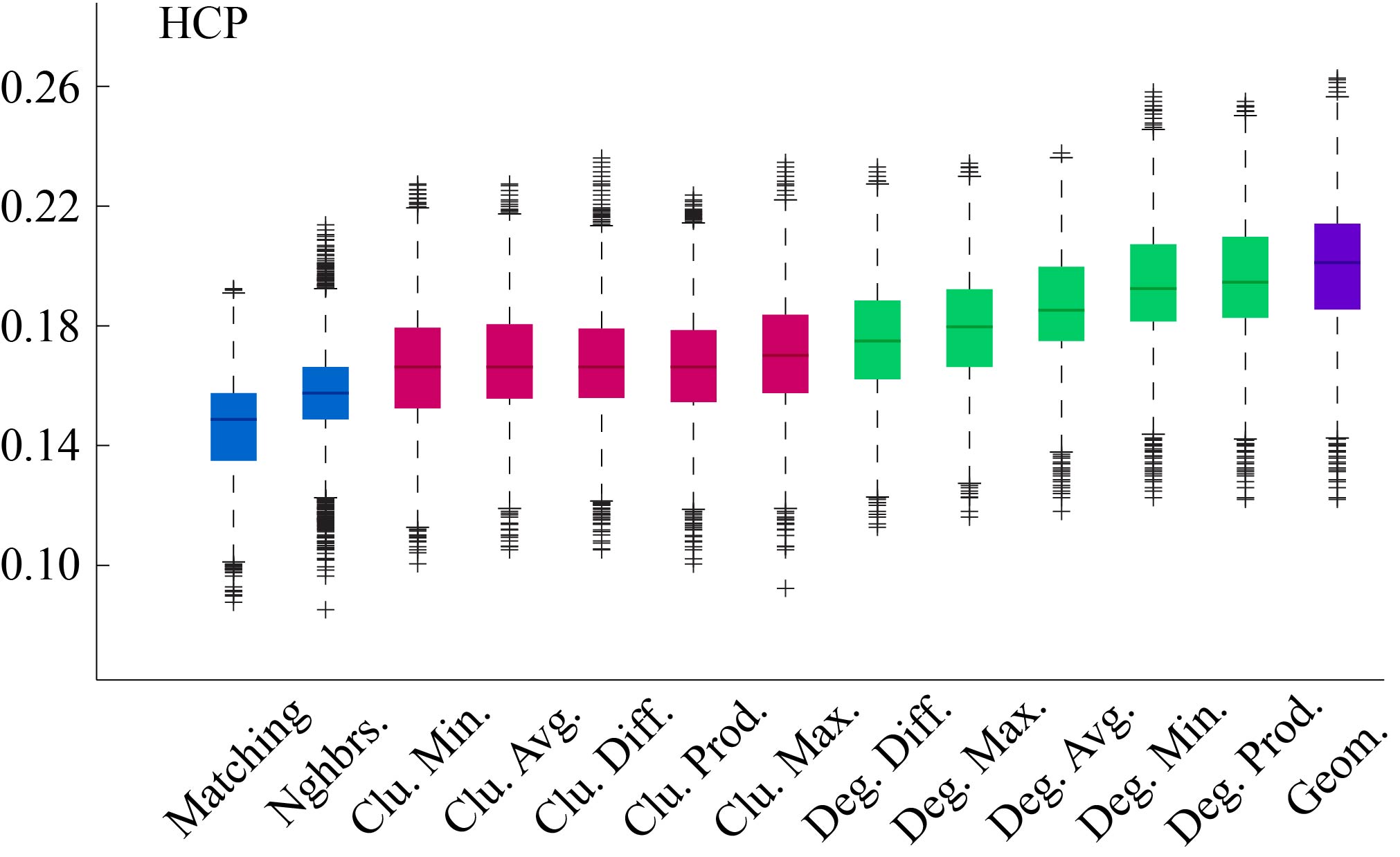}
\caption{Model energies for HCP dataset.}
\label{fig:figsi1}
\end{figure}

\begin{figure}[ht]
\centering
\includegraphics[width=\linewidth]{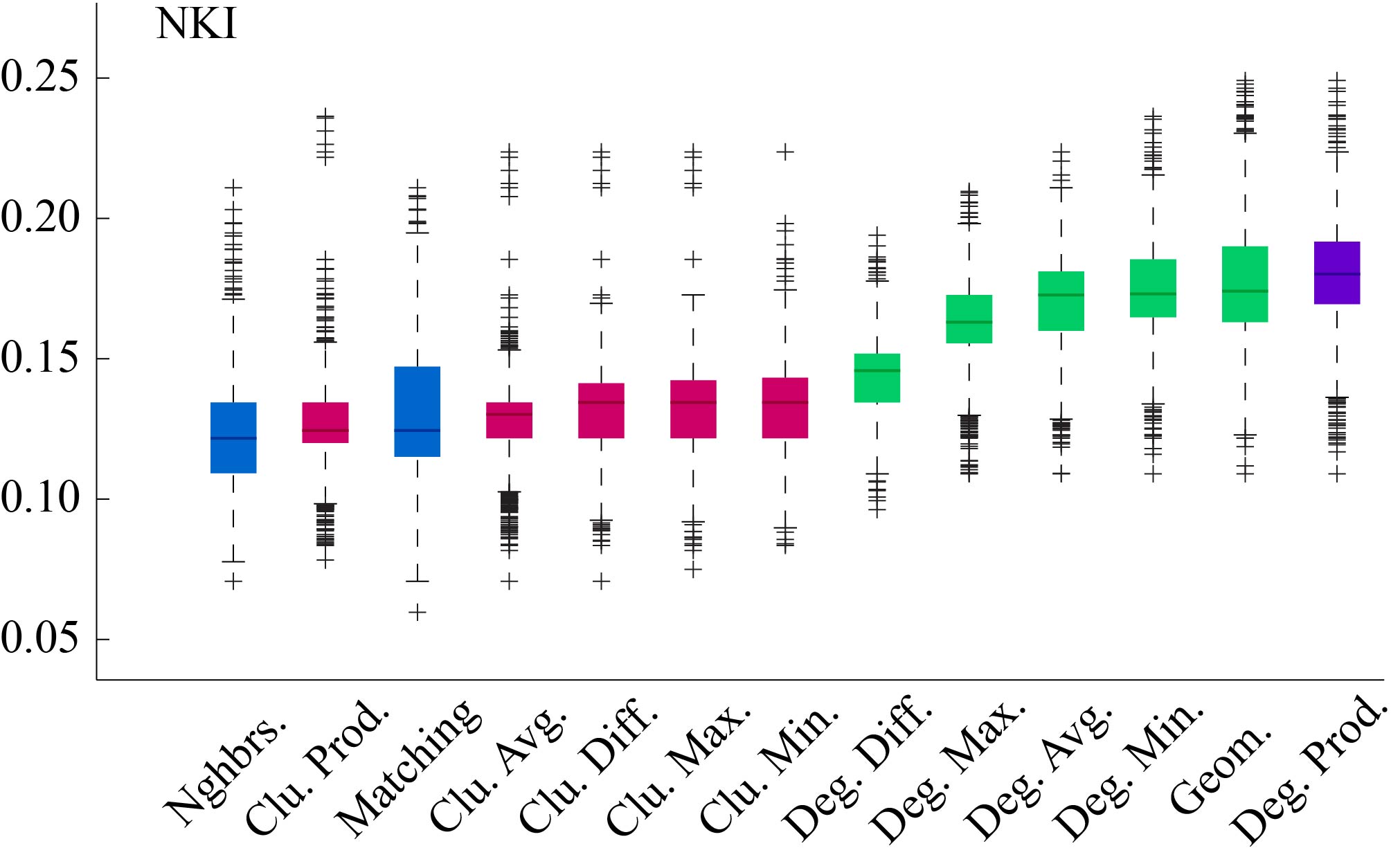}
\caption{Model energies for NKI dataset.}
\label{fig:figsi2}
\end{figure}

\begin{figure}[ht]
\centering
\includegraphics[width=\linewidth]{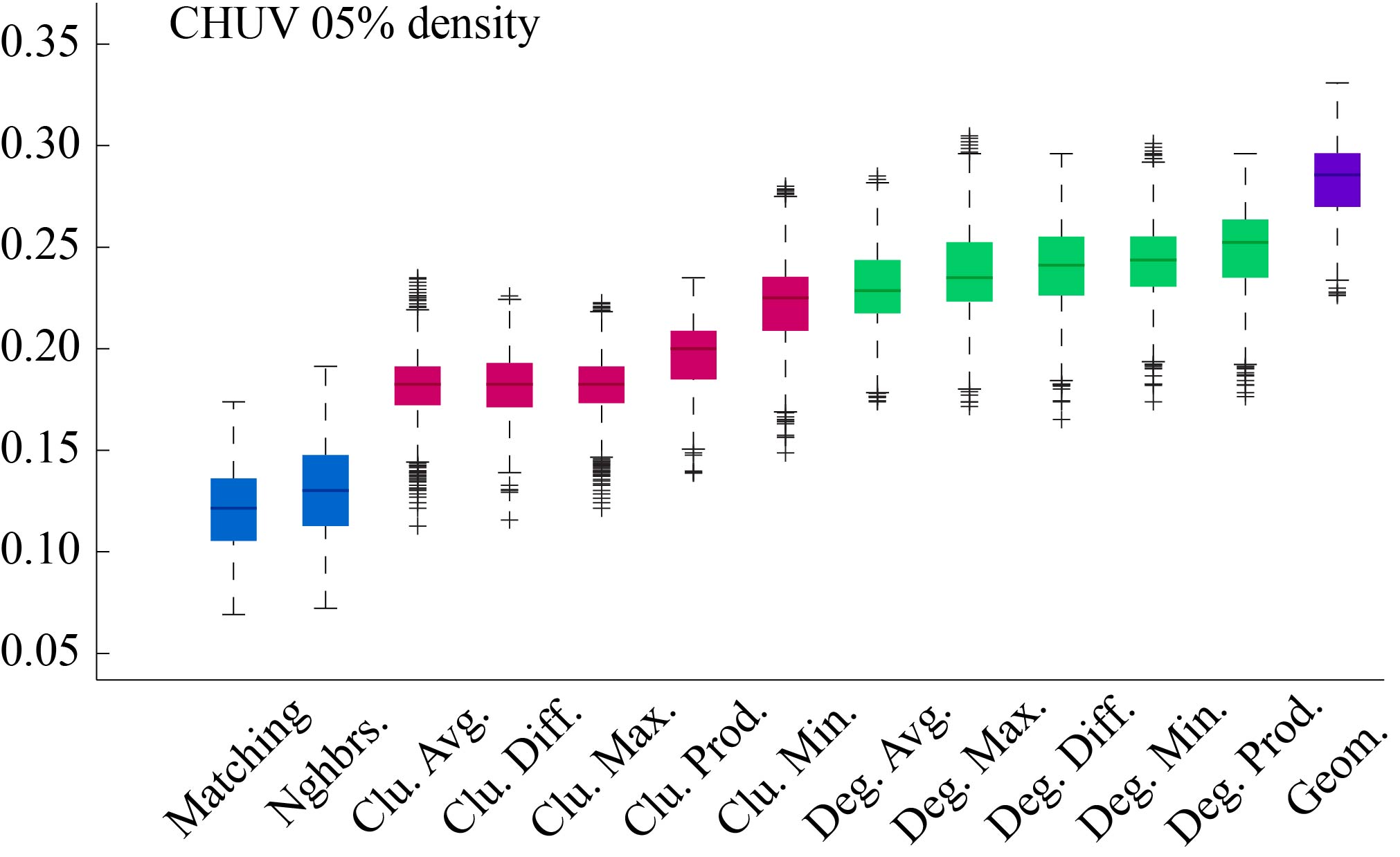}
\caption{Model energies for CHUV dataset with $\rho\approx5\%$.}
\label{fig:figsi3}
\end{figure}

\begin{figure}[ht]
\centering
\includegraphics[width=\linewidth]{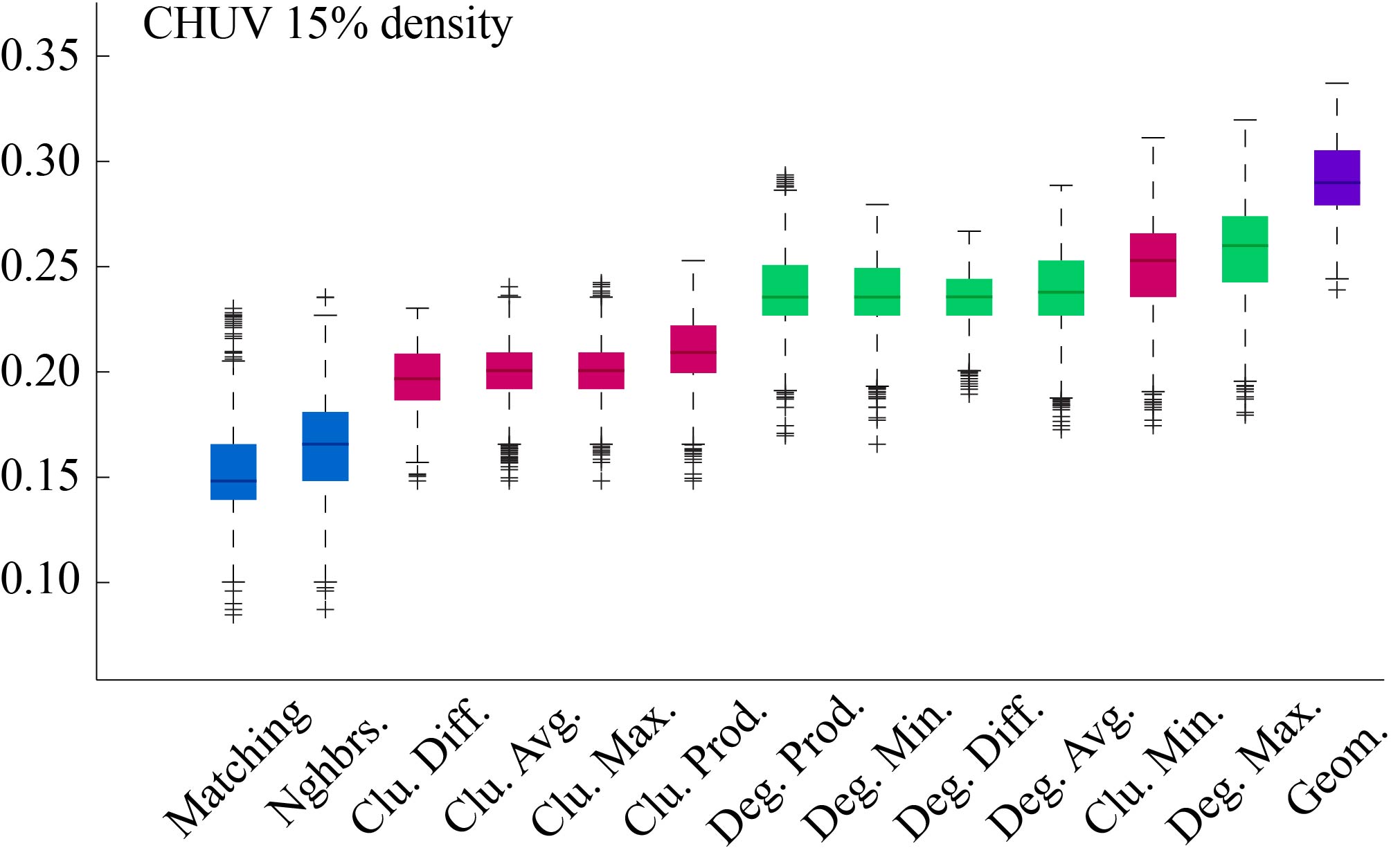}
\caption{Model energies for CHUV dataset with $\rho\approx15\%$.}
\label{fig:figsi4}
\end{figure}

\begin{figure}[ht]
\centering
\includegraphics[width=\linewidth]{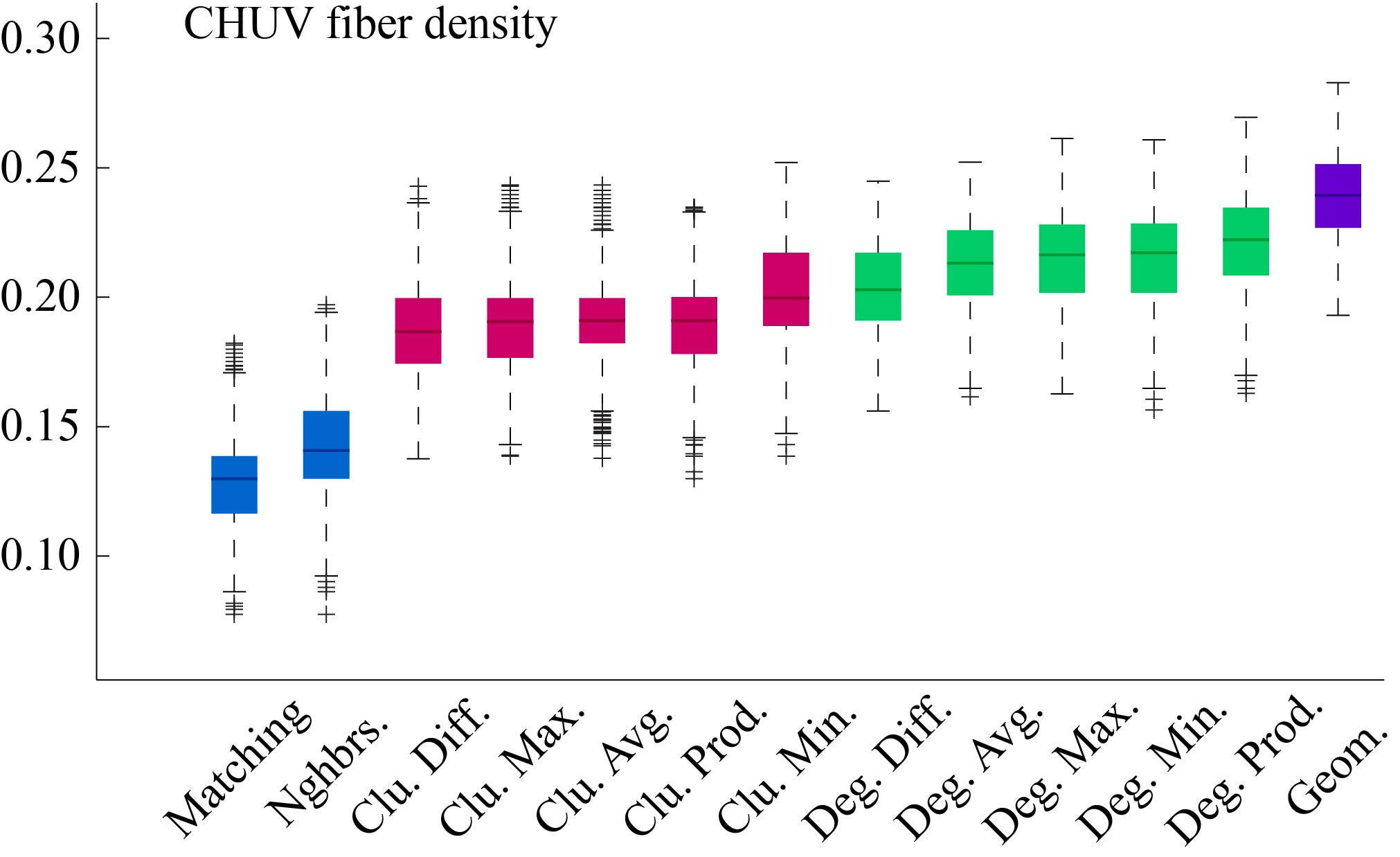}
\caption{Model energies for CHUV dataset with $\rho\approx10\%$ and edge presence/absence  determined by fiber density weights rather than streamline/fiber tract counts.}
\label{fig:figsi5}
\end{figure}

\begin{figure}[ht]
\centering
\includegraphics[width=\linewidth]{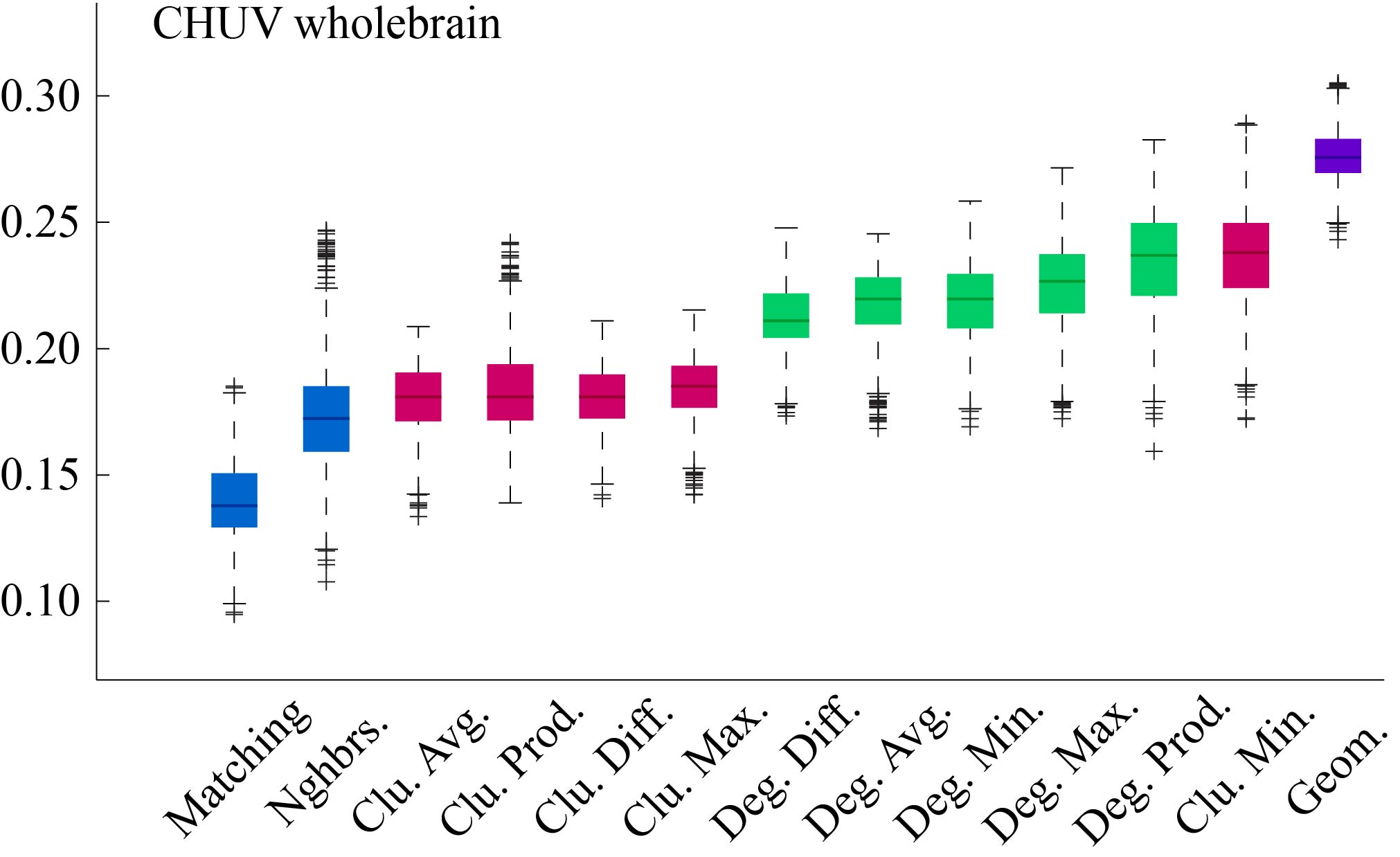}
\caption{Model energies for CHUV dataset with $\rho\approx10\%$ but for entire cerebral cortex.}
\label{fig:figsi6}
\end{figure}

\begin{figure}[ht]
\centering
\includegraphics[width=\linewidth]{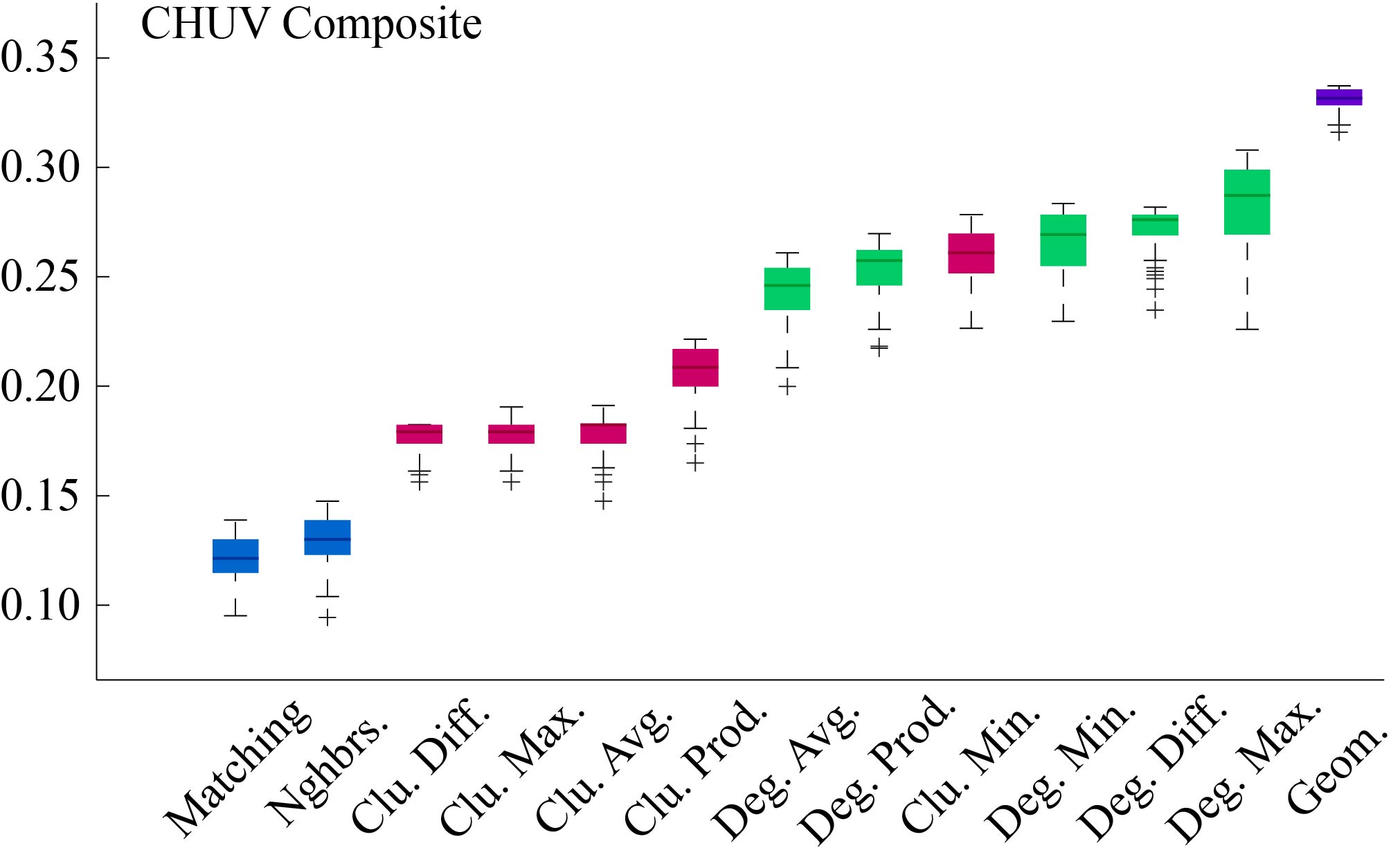}
\caption{Model energies for CHUV composite connectivity matrix.}
\label{fig:figsi7}
\end{figure}

\begin{figure}[ht]
\centering
\includegraphics[width=\linewidth]{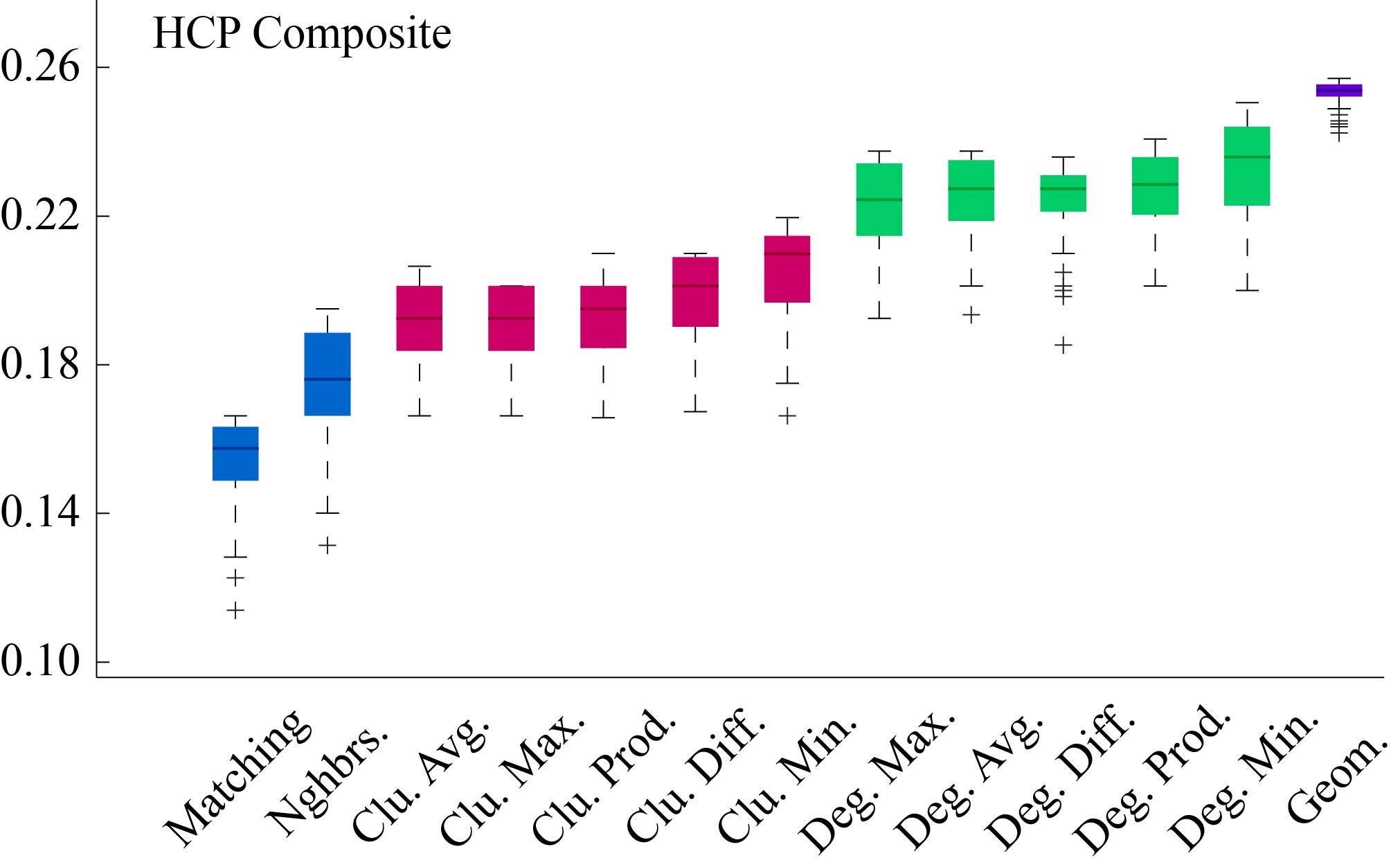}
\caption{Model energies for HCP composite connectivity matrix.}
\label{fig:figsi8}
\end{figure}

\begin{figure}[ht]
\centering
\includegraphics[width=\linewidth]{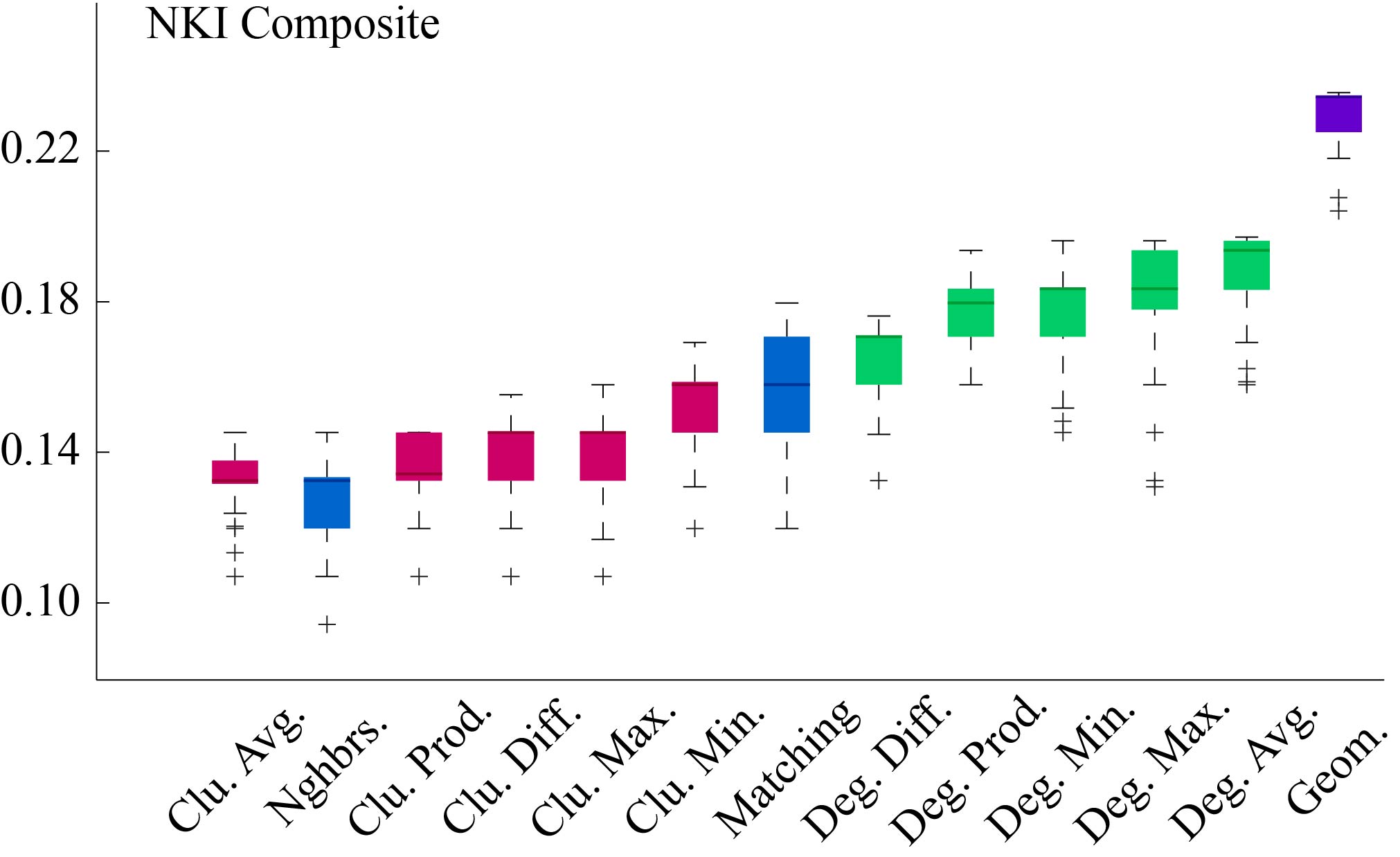}
\caption{Model energies for NKI composite connectivity matrix.}
\label{fig:figsi9}
\end{figure}

\begin{figure}[ht]
\centering
\includegraphics[width=\linewidth]{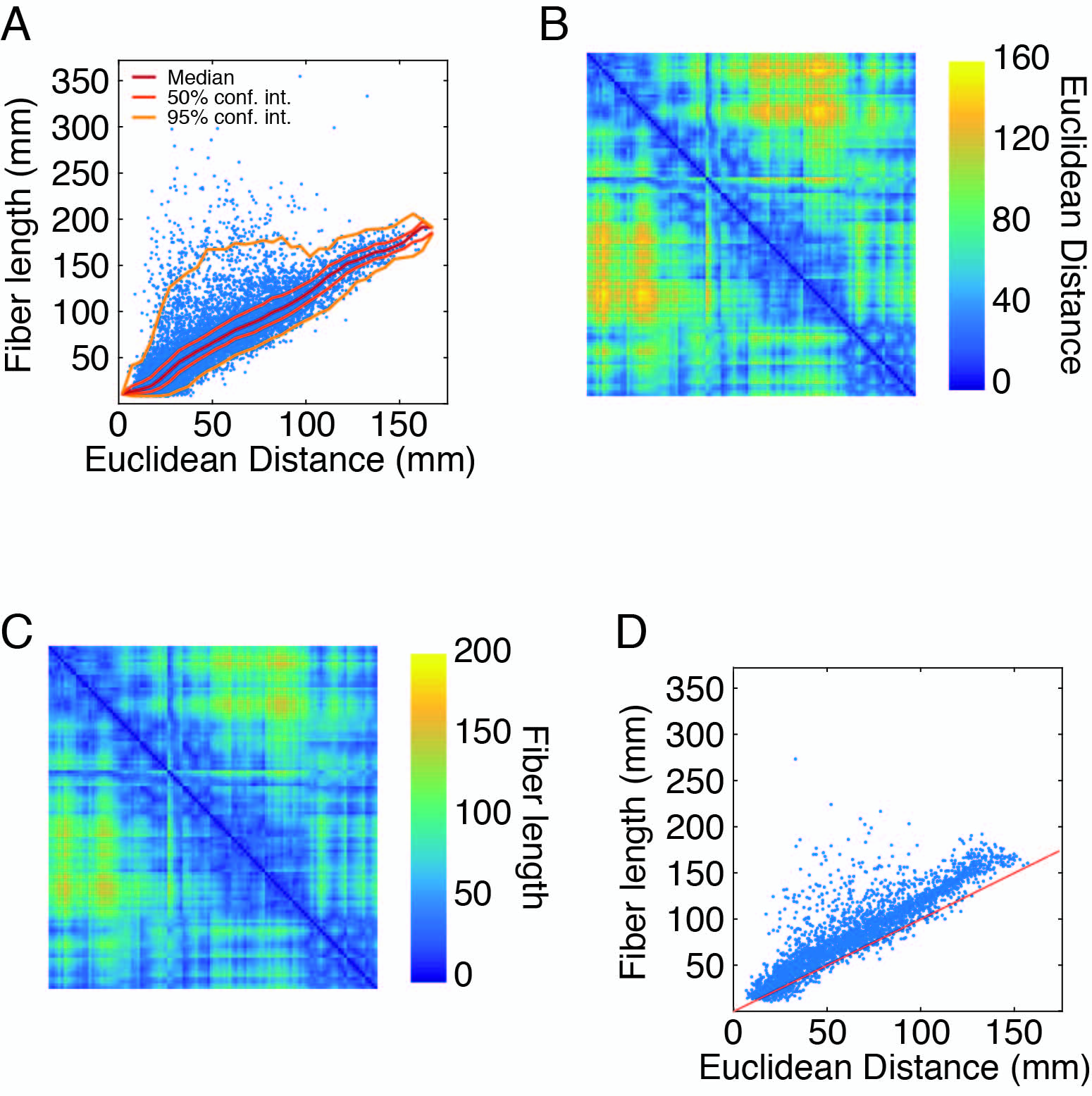}
\caption{Relationship between Euclidean distance and fiber length. (A) Scatter plot of Euclidean distance and fiber length for connections across all participants in the CHUV dataset. (B) The group-average Euclidean distance matrix obtained as the average distance between centroids across all participants. (C) Interpolated fiber length matrix, generated by the procedure described in the Appendix. (D) Scatter plot of group-average Euclidean distance versus interpolated fiber length. The red line in panel D is the identity line. Note that most connections have longer fiber length than Euclidean distance. However, a small number of connections have shorter fiber lengths. This occurs for connections that originate and terminate near the boundary of two parcels. In this case, the fiber length can be very short, while the Euclidean distance between the parcels can be long.}
\label{fig:figsi10}
\end{figure}

\begin{figure}[ht]
\centering
\includegraphics[width=\linewidth]{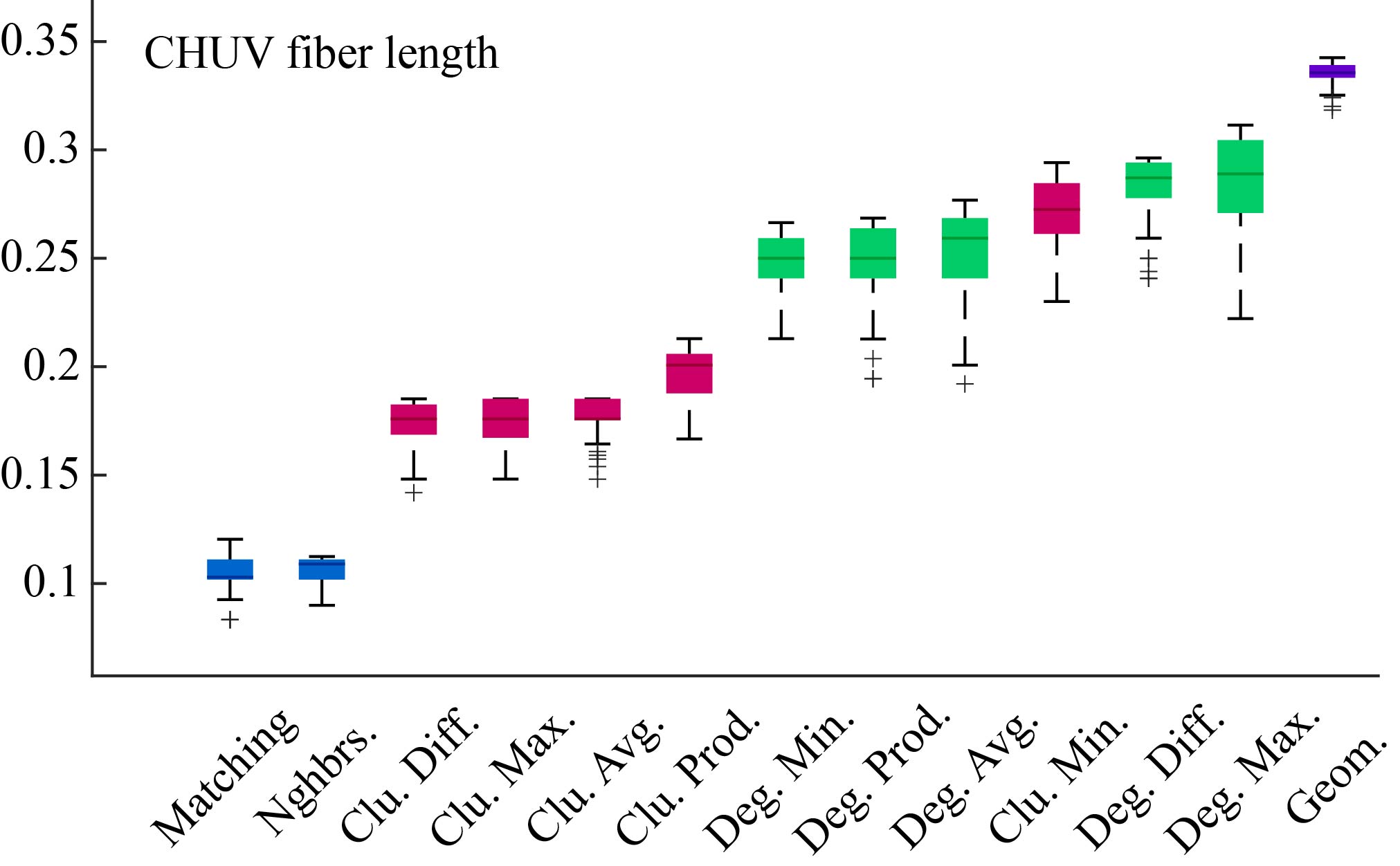}
\caption{Model energies for CHUV composite connectivity matrix using the interpolated fiber length matrix in place of Euclidean distance.}
\label{fig:figsi11}
\end{figure}

\begin{figure}[ht]
\centering
\includegraphics[width=\linewidth]{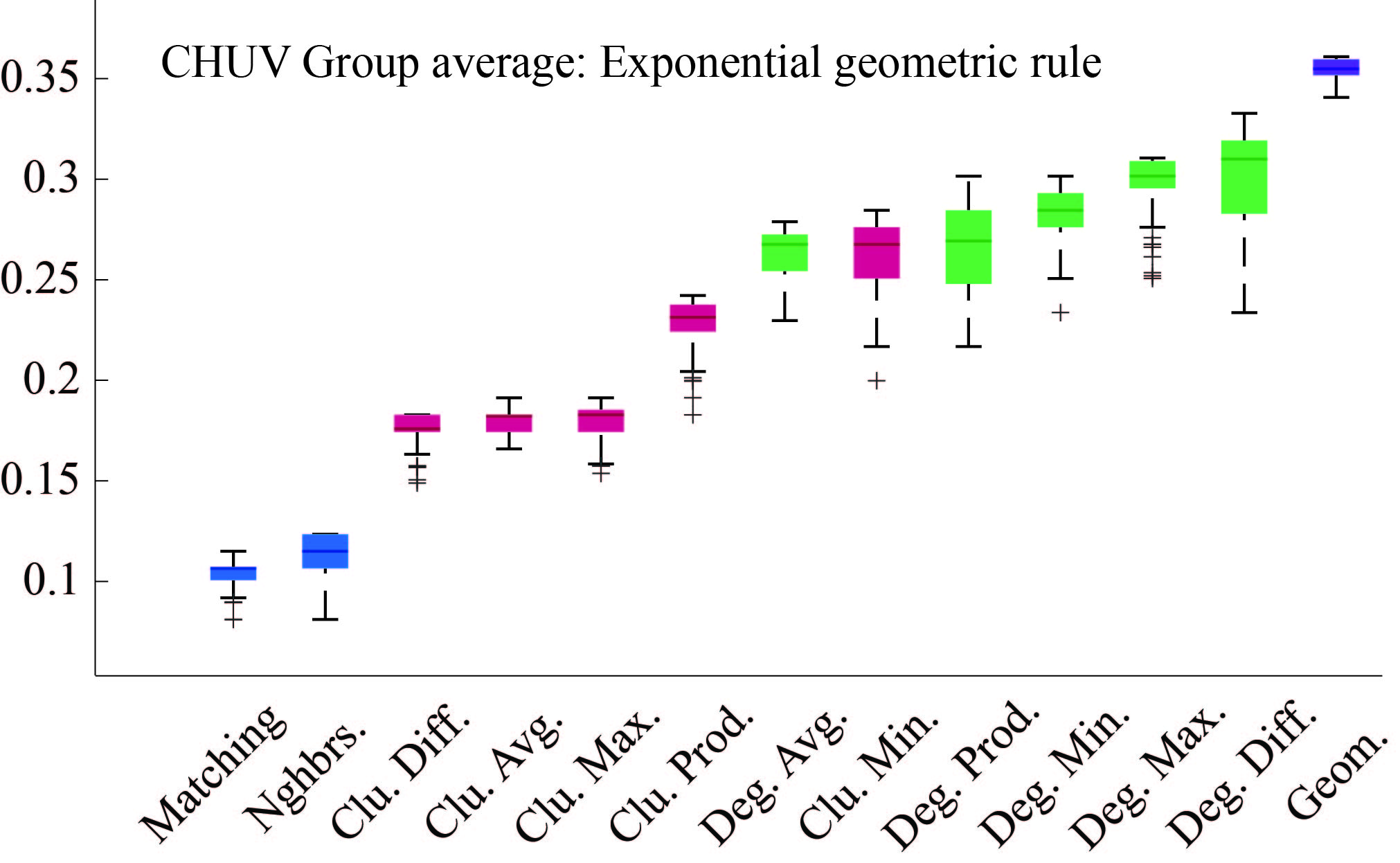}
\caption{Model energies for CHUV composite connectivity matrix using an exponential function in place of the power-law function for the geometric wiring rule.}
\label{fig:figsi12}
\end{figure}

\clearpage

\begin{figure}[ht]
\centering
\includegraphics[width=\linewidth]{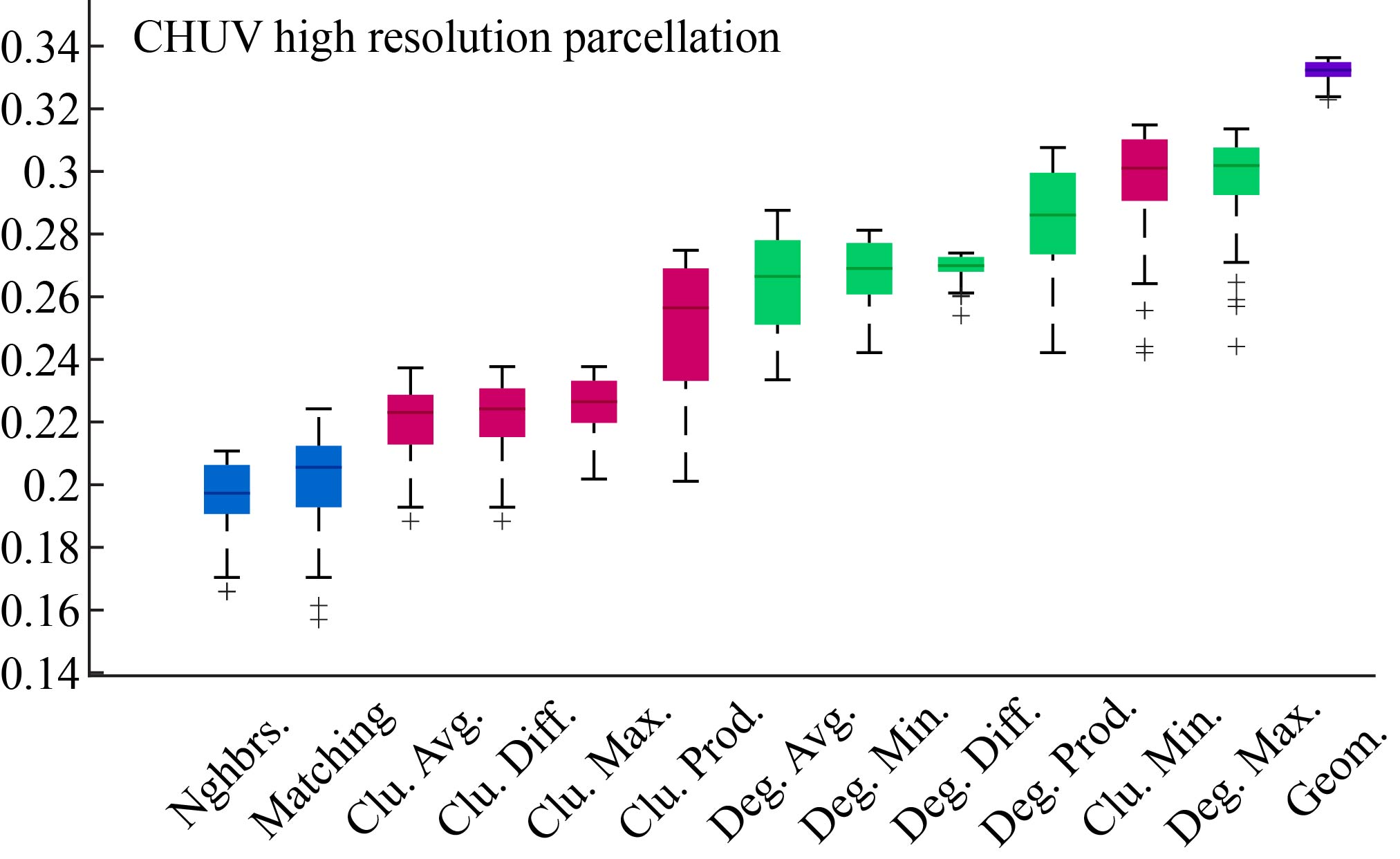}
\caption{Model energies for CHUV composite connectivity matrix with a higher-resolution partition ($n=455$ cortical nodes; $n=223$ cortical nodes in the right hemisphere).}
\label{fig:figsi13}
\end{figure}

\begin{figure}[ht]
\centering
\includegraphics[width=\linewidth]{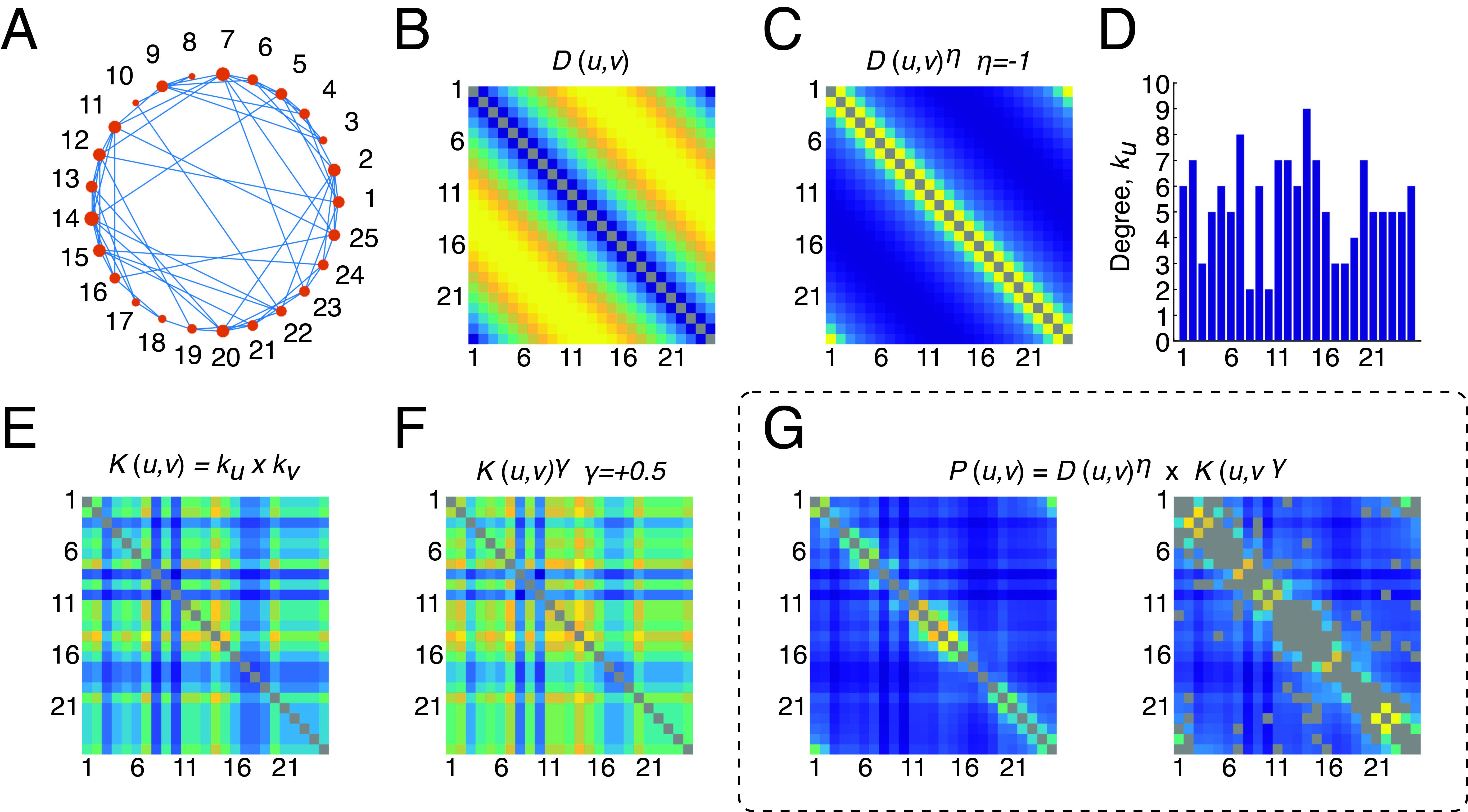}
\caption{Detailed explanation of how the generative model works. Panel A shows a toy network, with nodes embedded along the perimeter of the unit circle. Given this network and our generative models, we can ask the following question: If we wish to add a new edge to the network according to the wiring rule: $P(u,v) = K(u,v)^\gamma \times D(u,v)^\eta$, where will that edge most likely go? To answer this question, we need to first calculate the distance between all pairs of nodes (panel B), whose elements we raise to the power $\eta = -1$ (panel C). The other component we need is the matrix, $K(u,v)$, which represents the non-geometric component. One possible definition of $K(u,v)$ is the product of node degrees (the \textit{deg. prod} model). Given this particular definition, we set $K(u,v) = k_u \times k_v$. To generate this matrix, we first calculate $k_u$ for all $u$ (panel D). Then we multiply $k_u \times k_v$ for all pairs of nodes, $\{u,v\}$ (panel E). We next raise $K(u,v)$ to the power $\gamma = 1$, which we show in panel F. We perform the element-wise multiplication of $K(u,v)^\gamma$ by $D(u,v)^\eta$ (panel G, left). Finally, we have to remove the pairs, $\{u,v\}$, for which a connection already exists (the gray cells in panel G, right). The nonzero elements of this matrix give us the relative probabilities of where an edge would get placed given this wiring rule. After placing the edge according to these probability, the model would return to panel A and the process would repeat itself.}
\label{fig:figsi14}
\end{figure}

\begin{figure}[ht]
\centering
\includegraphics[width=\linewidth]{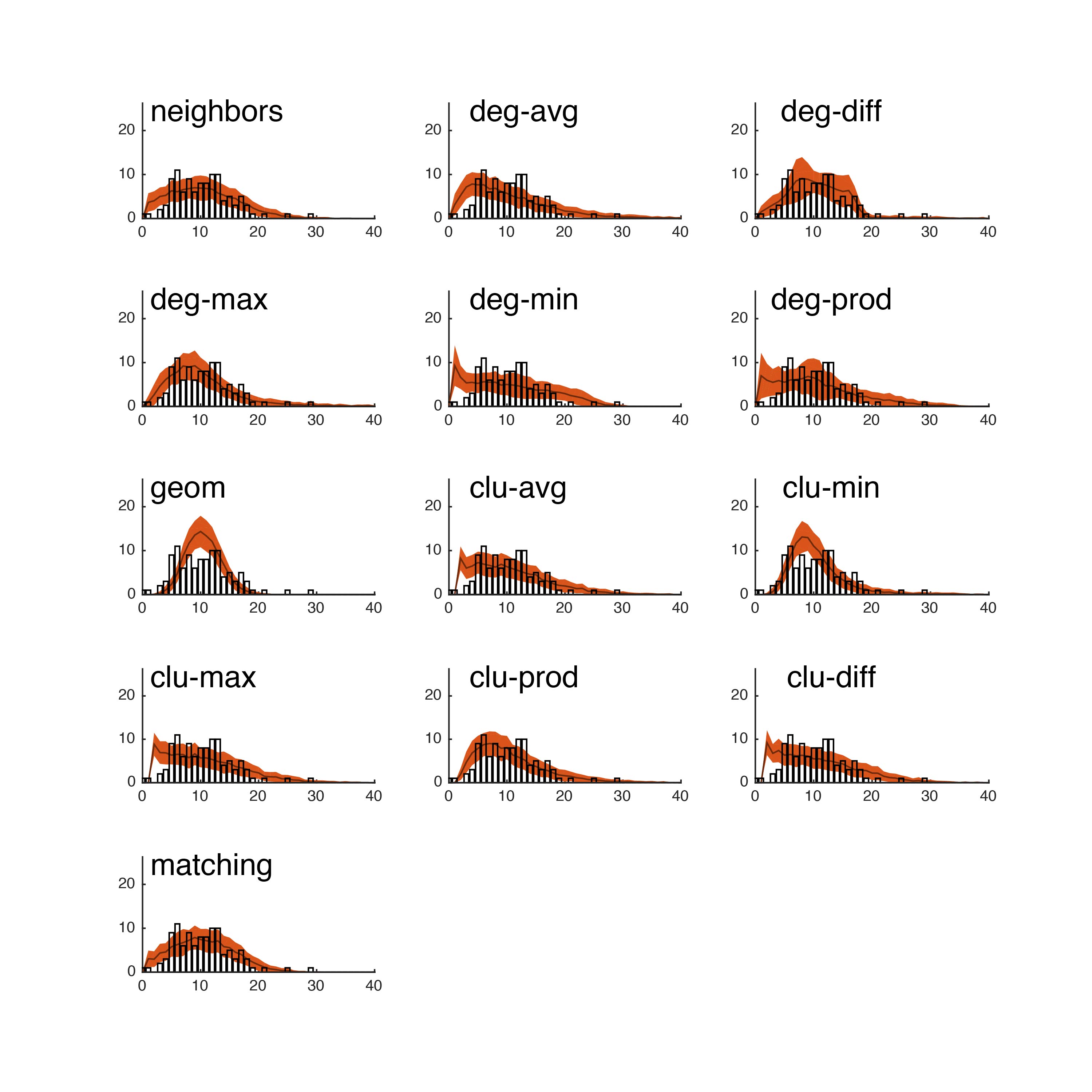}
\caption{We show the observed degree distribution (black bar plot) against the degree distributions of the best-fitting synthetic networks generated with each model. In all panels, the x-axis indicates degree ($k$) and the y-axis indicates frequency. This figure shows data from a single representative subject in the CHUV cohort.}
\label{fig:figsi15}
\end{figure}

\begin{figure}[ht]
\centering
\includegraphics[width=\linewidth]{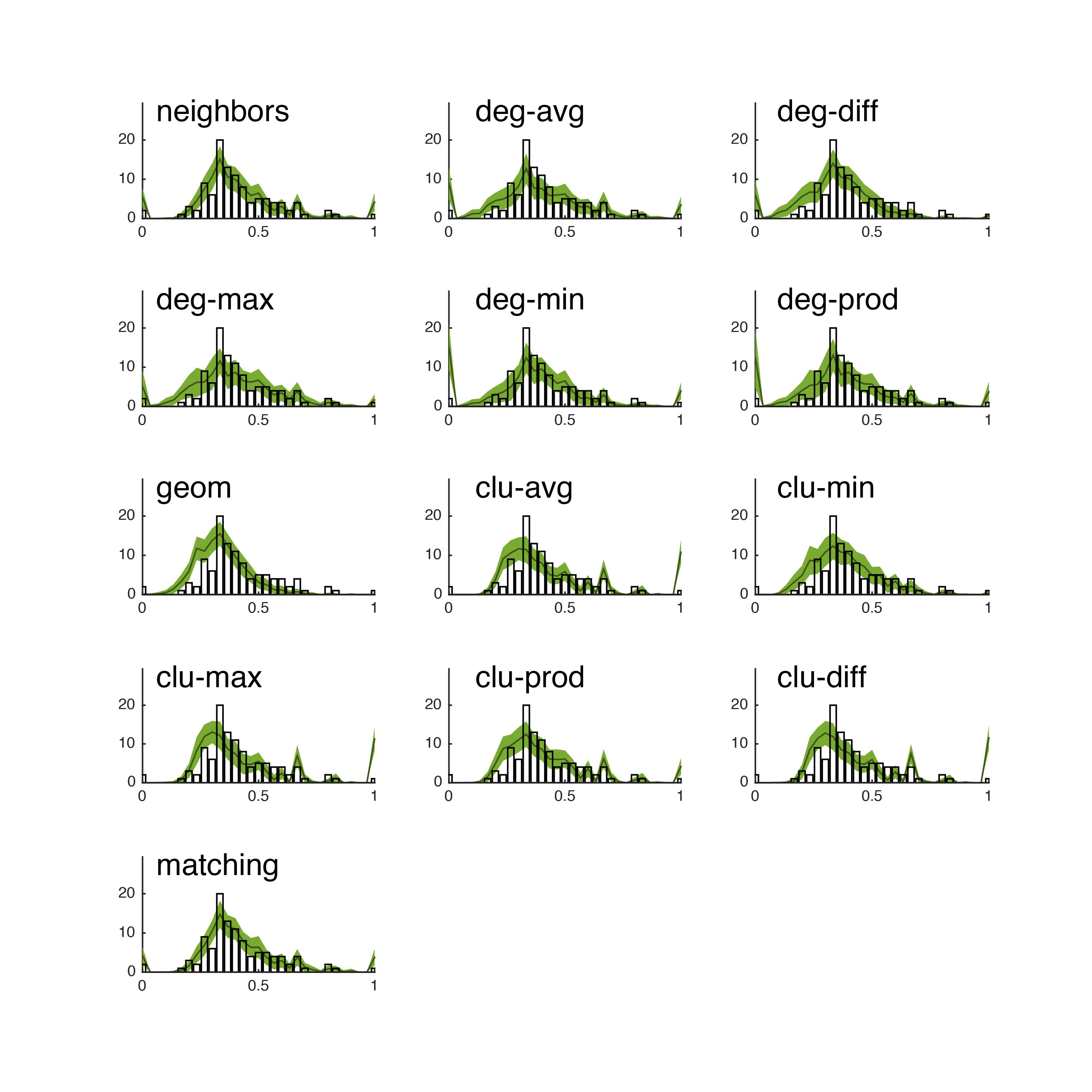}
\caption{We show the observed clustering coefficient distribution (black bar plot) against the clustering coefficient distributions of the best-fitting synthetic networks generated with each model. In all panels, the x-axis indicates clustering coefficient ($c$) and the y-axis indicates frequency. This figure shows data from a single representative subject in the CHUV cohort.}
\label{fig:figsi16}
\end{figure}

\begin{figure}[ht]
\centering
\includegraphics[width=\linewidth]{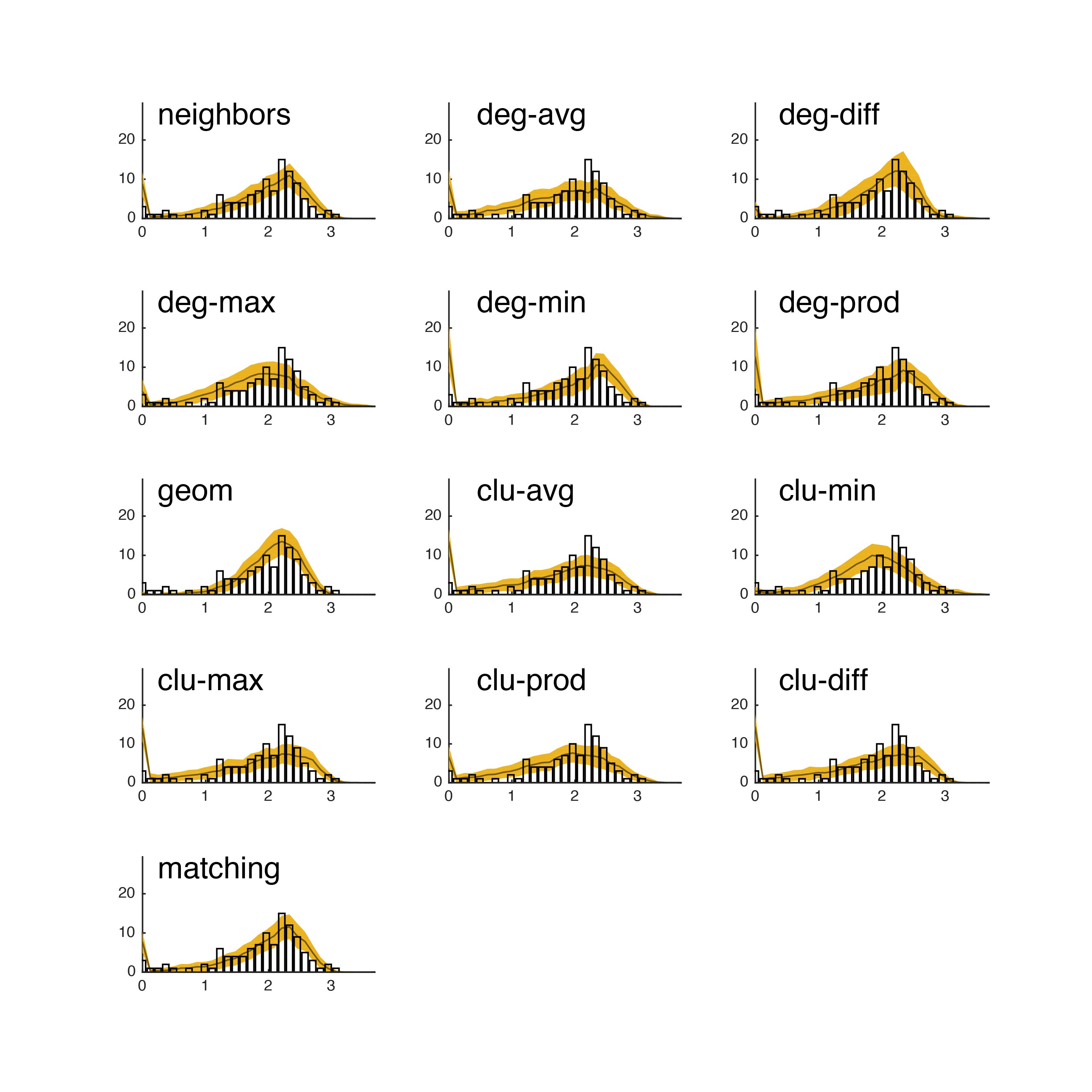}
\caption{We show the observed betweenness centrality distribution (black bar plot) against the betweenness centrality distributions of the best-fitting synthetic networks generated with each model. In all panels, the x-axis indicates betweenness centrality ($b$) and the y-axis indicates frequency. It should be noted, that due the orders of magnitude difference between the most central and least central nodes, the x-axis has been log-transformed. This figure shows data from a single representative subject in the CHUV cohort.}
\label{fig:figsi17}
\end{figure}

\begin{figure}[ht]
\centering
\includegraphics[width=\linewidth]{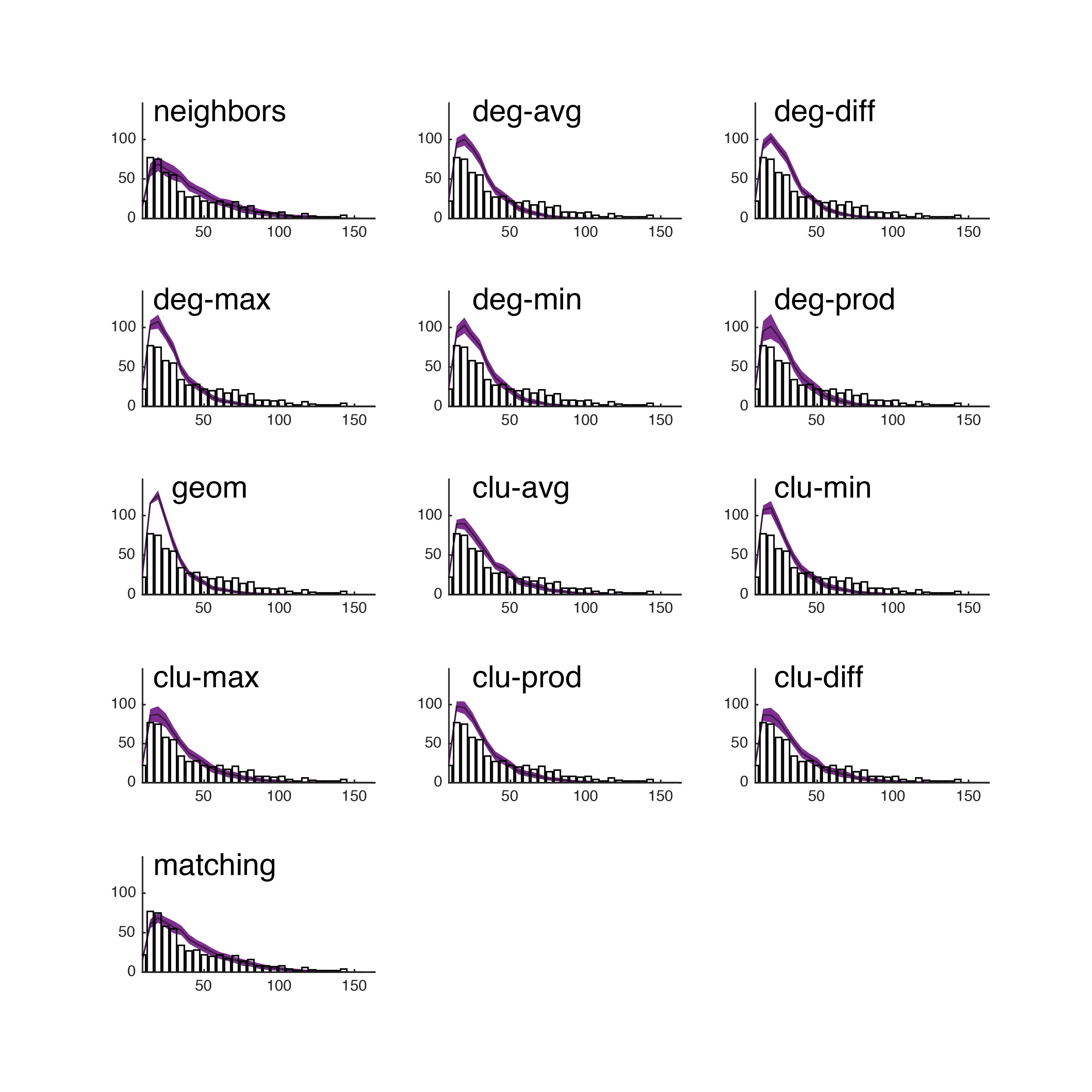}
\caption{We show the observed edge length distribution (black bar plot) against the edge length distributions of the best-fitting synthetic networks generated with each model. In all panels, the x-axis indicates edge length ($e$) and the y-axis indicates frequency. This figure shows data from a single representative subject in the CHUV cohort. Note that of the models shown here, "matching" and "neighbors" do a reasonable job generating networks with a broad range of connection lengths. This is in stark contrast with the geometric model "geom," which over-represents short range connections.}
\label{fig:figsi18}
\end{figure}
\clearpage

\begin{figure}[ht]
\centering
\includegraphics[width=\linewidth]{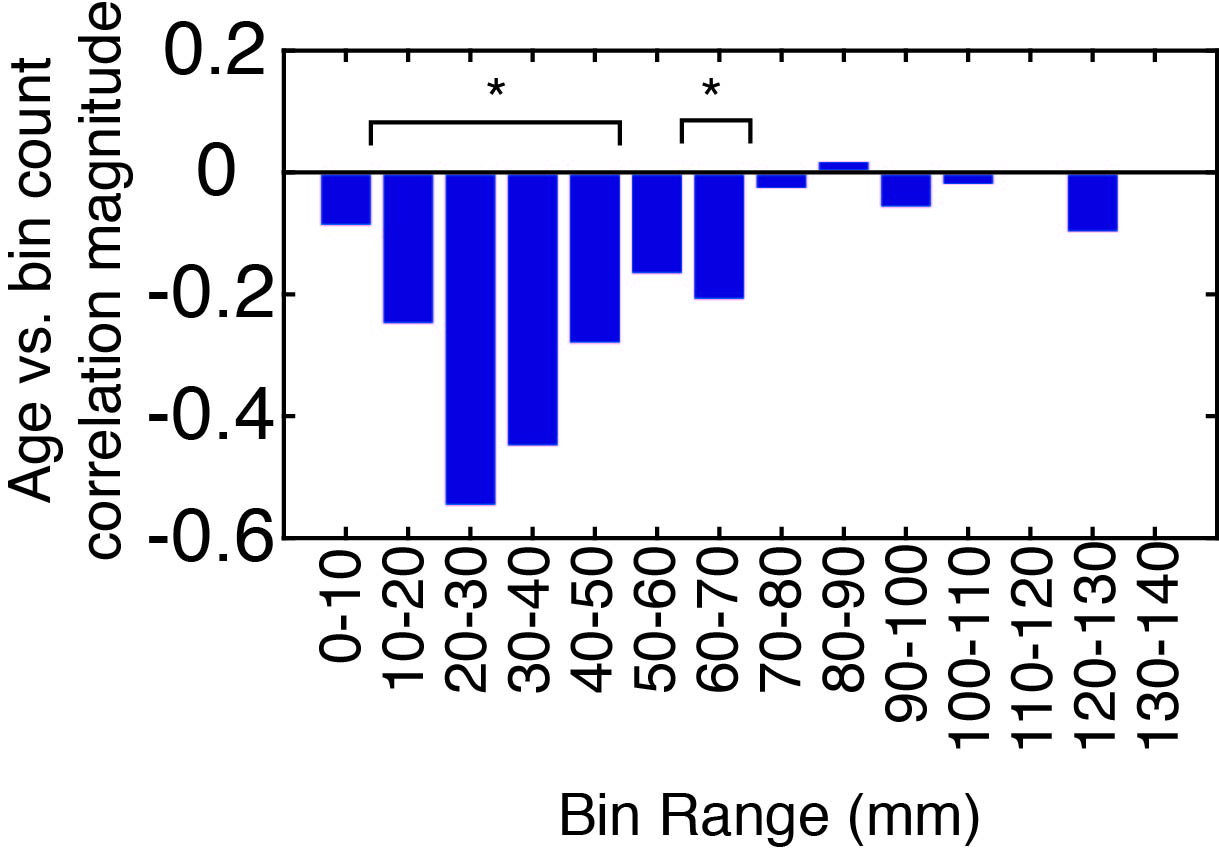}
\caption{Correlation of age with number of connections of particular lengths. The height of each bar represents the magnitude of the Pearson's correlation coefficient. The bins with an asterisk exhibit statistically significant change (p < 0.05, uncorrected).}
\label{fig:figsi19}
\end{figure}
\clearpage

\begin{sidewaystable}
\centering
\begin{tabular}{|l|l|l|l|l|l|l|l|l|}
\hline
Name & $K(u,v)$ & $E$ & $\eta$ & $\gamma$ & $KS_k$ & $KS_b$ & $KS_e$ & $KS_c$ \\
\hline
Clu. Avg. & $(\frac{c_u}{2} + \frac{c_v}{2})$ & $0.19 \pm 0.02$ & $-3.06 \pm 0.48$ & $-5.75 \pm 1.62$ & $0.14 \pm 0.03$ & $0.18 \pm 0.03$ & $0.17 \pm 0.03$ & $0.15 \pm 0.03$ \\
\hline
Clu. Diff. & $|c_u - c_v|$ & $0.19 \pm 0.02$ & $-3.13 \pm 0.49$ & $-5.85 \pm 2.07$ & $0.15 \pm 0.03$ & $0.18 \pm 0.03$ & $0.17 \pm 0.03$ & $0.16 \pm 0.03$ \\
\hline
Clu. Max. & $max [c_u,c_v]$ & $0.19 \pm 0.02$ & $-3.14 \pm 0.49$ & $-5.75 \pm 1.97$ & $0.15 \pm 0.03$ & $0.18 \pm 0.03$ & $0.17 \pm 0.03$ & $0.15 \pm 0.03$ \\
\hline
Clu. Min & $min [c_u,c_v]$ & $0.24 \pm 0.03$ & $-3.60 \pm 0.41$ & $-4.04 \pm 1.09$ & $0.12 \pm 0.03$ & $0.19 \pm 0.05$ & $0.23 \pm 0.03$ & $0.17 \pm 0.05$ \\
\hline
Clu. Prod. & $c_uc_v$ & $0.20 \pm 0.02$ & $-3.14 \pm 0.39$ & $-3.44 \pm 1.00$ & $0.11 \pm 0.03$ & $0.18 \pm 0.03$ & $0.19 \pm 0.03$ & $0.15 \pm 0.04$ \\
\hline
Deg. Avg. & $(\frac{k_u}{2} + \frac{k_v}{2})$ & $0.24 \pm 0.02$ & $-3.76 \pm 0.57$ & $2.47 \pm 0.38$ & $0.18 \pm 0.05$ & $0.18 \pm 0.05$ & $0.23 \pm 0.03$ & $0.19 \pm 0.04$ \\
\hline
Deg. Diff. &  $|k_u - k_v|$ & $0.25 \pm 0.02$ & $-4.55 \pm 0.76$ & $-1.03 \pm 2.61$ & $0.13 \pm 0.04$ & $0.19 \pm 0.04$ & $0.24 \pm 0.03$ & $0.21 \pm 0.05$ \\
\hline
Deg. Max. & $max [k_u,k_v]$ & $0.25 \pm 0.02$ & $-4.02 \pm 0.60$ & $2.20 \pm 0.43$ & $0.13 \pm 0.04$ & $0.18 \pm 0.05$ & $0.24 \pm 0.03$ & $0.18 \pm 0.05$ \\
\hline
Deg. Min. & $min [k_u,k_v]$ & $0.25 \pm 0.02$ & $-3.99 \pm 0.64$ & $2.03 \pm 0.47$ & $0.20 \pm 0.05$ & $0.16 \pm 0.04$ & $0.24 \pm 0.03$ & $0.21 \pm 0.04$ \\
\hline
Deg. Prod. & $k_uk_v$ & $0.26 \pm 0.02$ & $-3.83 \pm 0.70$ & $1.14 \pm 0.32$ & $0.19 \pm 0.07$ & $0.16 \pm 0.06$ & $0.24 \pm 0.04$ & $0.22 \pm 0.04$ \\
\hline
Matching & $\frac{|\mathbf{x}_{u \setminus v}\cap\mathbf{x}_{v \setminus u}|}{|\mathbf{x}_{u \setminus v}\cup\mathbf{x}_{v \setminus u}|}$ & $0.12 \pm 0.02$ & $-0.98 \pm 0.37$ & $0.42 \pm 0.04$ & $0.10 \pm 0.03$ & $0.10 \pm 0.02$ & $0.10 \pm 0.03$ & $0.11 \pm 0.02$ \\
\hline
Nghbrs. & $\sum_w a_{uw}a_{wv}$ & $0.14 \pm 0.02$ & $-1.18 \pm 0.43$ & $0.35 \pm 0.04$ & $0.11 \pm 0.03$ & $0.11 \pm 0.03$ & $0.11 \pm 0.03$ & $0.11 \pm 0.03$ \\
\hline
Geom. & $\mathbf{1}$ & $0.29 \pm 0.02$ & $-4.01 \pm 0.31$ & $N/A$ & $0.15 \pm 0.03$ & $0.18 \pm 0.04$ & $0.29 \pm 0.02$ & $0.27 \pm 0.03$ \\
\hline
\end{tabular}
\caption{\label{tab:tab1}Complete list of generative models. The first two columns show each model's name and the non-geometric wiring rule. The remaining columns indicate sample mean$\pm$standard error energy ($E$), and the four KS statistics, $KS_k$, $KS_b$, $KS_e$, and $KS_c$.}
\end{sidewaystable}

\end{document}